\documentclass[%
10pt,
aps,
amsmath,    
amssymb,    
reprint,    
twocolumn,
english,    
floatfix,   
groupedaddress,     
superscriptaddress,
twocolumn, %
showpacs, %
showkeys,
]{revtex4-1}

\usepackage{amsmath}

\usepackage{amsfonts}

\usepackage{graphicx}
\usepackage[rgb]{xcolor}

\definecolor{mygrey}{gray}{0.35}
\definecolor{myblue}{rgb}{0.2,0.2,0.8}
\definecolor{myzard}{cmyk}{0,0,0.05,0}
\definecolor{mywhite}{rgb}{1,1,1}
\definecolor{myred}{rgb}{1,0.,0.3}

\usepackage{textcomp}

\usepackage[T1]{fontenc}	

\usepackage{lmodern}

\usepackage[utf8]{inputenc} 

\usepackage[english]{babel}

\usepackage{physics}

\usepackage[
    unicode=true,           
    colorlinks=true,
    citecolor=myblue,
    linkcolor=myblue,
    urlcolor=myblue,
    ]{hyperref}

\hypersetup{
    pdftitle={Clock transition by continuous dynamical decoupling of a three-level system}, 
    pdfauthor={Alexander Stark, Nati Aharon, Alexander Huck, Haitham A. R. El-Ella, Alex Retzker, Fedor Jelezko, Ulrik Lund Andersen}, 
    pdfsubject={Publication manuscript}, 
    pdfcreator={ShareLatex and TeXstudio 2.12.8}, 
    pdfproducer={pdfLaTeX, biber, and hyperref-package. No animals, MS-EULA or BSA-rules were harmed.}, 
    pdfkeywords={nitrogen-vacancy, three-level system, continuous dynamical decoupling, diamond, robust qubit}
}

\usepackage[nameinlink]{cleveref}

\crefname{figure}{Fig.}{Figs}
\crefname{equation}{eq.}{eqs}

\Crefname{figure}{Fig.}{Figs}
\Crefname{equation}{Eq.}{Eqs}

\begin{document}


\title[Manuscript File]{Clock transition by continuous dynamical decoupling of a three-level system}

\author{A. Stark$^{\dagger}$}
\thanks{}
\email[Email of corresponding author: ]{astark@fysik.dtu.dk}
\affiliation{Department of Physics, Technical University of Denmark, Fysikvej, Kongens Lyngby 2800, Denmark}
\affiliation{Institute for Quantum Optics, Ulm University, Albert-Einstein-Allee 11, Ulm 89081, Germany}

\author{N. Aharon}
\thanks{These authors contributed equally}
\affiliation{Racah Institute of Physics, The Hebrew University of Jerusalem, Jerusalem 91904, Israel}

\author{A. Huck}
\affiliation{Department of Physics, Technical University of Denmark, Fysikvej, Kongens Lyngby 2800, Denmark}

\author{H.A.R. El-Ella}
\affiliation{Department of Physics, Technical University of Denmark, Fysikvej, Kongens Lyngby 2800, Denmark}

\author{A. Retzker}
\affiliation{Racah Institute of Physics, The Hebrew University of Jerusalem, Jerusalem 91904, Israel}

\author{F. Jelezko}
\affiliation{Institute for Quantum Optics, Ulm University, Albert-Einstein-Allee 11, Ulm 89081, Germany}
\affiliation{Center for Integrated Quantum Science and Technology (IQ$^\text{{st}}$), Ulm University, 89081 Germany}

\author{U.L. Andersen}
\affiliation{Department of Physics, Technical University of Denmark, Fysikvej, Kongens Lyngby 2800, Denmark}

\begin{abstract}
    We present a novel continuous dynamical decoupling scheme for the construction of a robust qubit in a three-level system. 
    By means of a clock transition adjustment, we first show how robustness to environmental noise is achieved, while eliminating drive-noise, to first-order. 
    We demonstrate this scheme with the spin sub-levels of the NV-centre’s electronic ground state.
    By applying drive fields with moderate Rabi frequencies, the drive noise is eliminated and an improvement of 2 orders of magnitude in the coherence time is obtained compared to the pure dephasing time. 
    We then show how the clock transition adjustment can be tuned to eliminate also the second-order effect of the environmental noise with moderate drive fields. 
    A further improvement of more than 1 order of magnitude in the coherence time is expected and confirmed by simulations.  
    Hence, our scheme prolongs the coherence time towards the lifetime-limit using a relatively simple experimental setup.
\end{abstract}



\maketitle

\section{Introduction}

The reliable and efficient construction and manipulation of qubits is necessary for the implementation of quantum technological applications and quantum information processing. 
In solid-state and atomic systems, ambient magnetic field fluctuations constitute a serious impediment, which usually limits the coherence time to a small fraction of the inherent lifetime. 
Pulsed dynamical decoupling \cite{hahn_spin_1950, carr_effects_1954, meiboom_modified_1958} has proven to be very useful in prolonging the coherence time \cite{viola_dynamical_1998, biercuk_optimized_2009, du_preserving_2009, lange_universal_2010, ryan_robust_2010, naydenov_dynamical_2011, wang_comparison_2012, bar-gill_solid-state_2013, baumgart_ultrasensitive_2016}. 
However, in order to mitigate both environmental and controller noise, composite high-frequency pulse sequences must usually be applied \cite{khodjasteh_fault-tolerant_2005, uhrig_keeping_2007, souza_robust_2011, yang_preserving_2011, farfurnik_optimizing_2015}, which are not easily combined with other coherent qubit operations and often require large field strengths \cite{gordon_optimal_2008}.
Continuous dynamical decoupling \cite{fanchini_continuously_2007, gordon_optimal_2008, rabl_strong_2009, clausen_bath-optimized_2010, bermudez_electron-mediated_2011, bermudez_robust_2012, cai_long-lived_2012, xu_coherence-protected_2012, golter_protecting_2014, trypogeorgos_synthetic_2018} offers another possibility of suppressing environmental noise, where diminishing the effect of the controller noise can be achieved by different approaches. 
In this context, a rotary echo scheme \cite{mkhitaryan_decay_2014, mkhitaryan_highly_2015} can be viewed as analogous to pulsed dynamical decoupling.
The concatenation of several on-resonance driving fields \cite{cai_robust_2012, cohen_multi-qubit_2015, teissier_hybrid_2017, cohen_continuous_2017, stark_narrow-bandwidth_2017} is another concept, but inherently connected to a reduction of the dressed energy gap, eventually limiting the performance of the scheme, and in particular reducing the qubit gate operation time. 

Multi-state systems allow for yet a different approach. 
By applying continuous driving fields on a multi-level structure, a fully robust qubit - a qubit that is robust to both external and controller noise - can be obtained \cite{timoney_quantum_2011,aharon_general_2013}.
However, these multi-state schemes, which utilise on-resonance driving fields, are not applicable to a three-level system. 
A protected qubit subspace within a three-level configuration can be realised by the application of off-resonant strong driving fields \cite{aharon_fully_2016}, making the experimental realisation challenging.

In this report we show how a fully robust qubit can be simply constructed by means of a clock transition adjustment\cite{trypogeorgos_synthetic_2018} using a three-level system. 
We start with a basic version of our scheme where both continuous on-resonant and off-resonant driving fields are utilised. 
The on-resonant driving fields result in robustness to environmental noise, whereas the off-resonant driving fields facilitate robustness against driving noise, which typically limits continuous dynamical decoupling schemes. 
Similar to clock states, which possess a transition that is insensitive to first-order magnetic shifts for a given magnetic field value, the off-resonant driving fields generate a transition that is insensitive to first-order shifts in the drive-field amplitudes. 
We demonstrate this scheme by utilising the ground state spin level of the nitrogen-vacancy centre (NV) in diamond. 
The states are addressed by a combination of four microwave fields, adjusted to the same Rabi frequency, $\Omega$, at different transition frequencies, $\omega_1$ and $\omega_2$, and with a detuning, $\Delta$, respectively.
The spin states of the NV centre are initialised and read out by a 532\,nm laser identifying spin dependent fluorescence \cite{gruber_scanning_1997, jelezko_single_2006, maze_nanoscale_2008, doherty_nitrogen-vacancy_2013}. 
We are operating in the vicinity of the excited state level anti-crossing \cite{jacques_dynamic_2009, ivady_theoretical_2015} at a magnetic field of 35.8\,mT. 
At this bias field the intrinsic nitrogen nuclear spin becomes polarised through optical pumping and does not contribute to the level structure. 
Our implementation demonstrates that, to first order, drive-noise is eliminated, and compared to the pure dephasing time an improvement of 2 orders of magnitude in the coherence time is obtained even for a moderate drive field strength.
Finally, we present an improved version of the scheme, where the clock transition adjustment is extended to also eliminate the second-order effect of the environmental noise. 
Confirmed by simulations, our analysis shows that with a moderate driving field strength a further improvement of more than 1 order of magnitude in the coherence time can be obtained. 
Hence, our scheme allows prolonging the coherence time towards the lifetime limit in a simple experimental setup and without requiring exceptionally strong drive fields. 
The protocol is applicable to both the optical and microwave domain, and hence to a variety of atomic and solid state systems, such as trapped ions, rare-earth ions, and defect centres.

\section{The basic scheme}

\begin{figure}[!b]
    \centering
    \includegraphics[width=1.0\columnwidth]{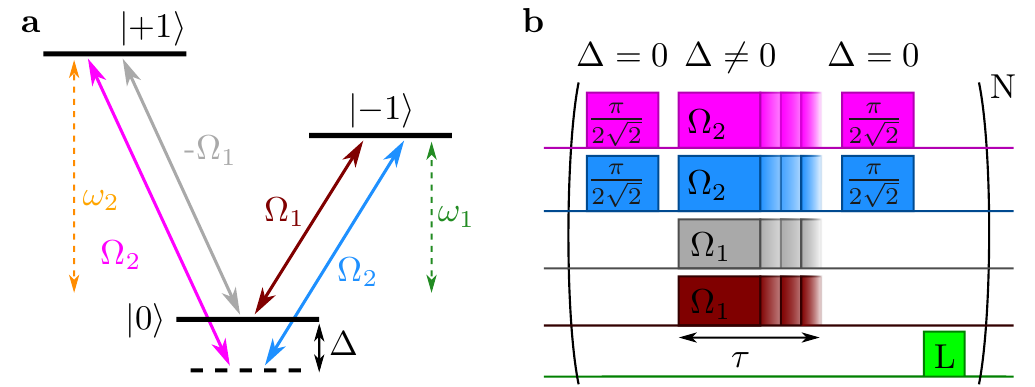}%
    \caption{Schematics for a double lambda ($\Lambda$) drive in the bare state representation. 
        (a) Four drives of strength, $\Omega_i$, form two (inverted) $\Lambda$ configurations.
        One lambda is driven resonantly, the other by a detuning of $\Delta$.
        Solid lines denote drives, whereas dotted lines indicate energy separations.
        (b) Schematic representation of the sequence applied in this work to measure the coherence times of the doubly-dressed states.
        The factor of $\sqrt{2}$ in the $\pi/2$-pulses reflect the $\Omega_B$ drive of the dressed states.
    }
    \label{fig:scheme}
\end{figure}

\begin{figure}[t]
    \centering
    \includegraphics[width=1.0\columnwidth]{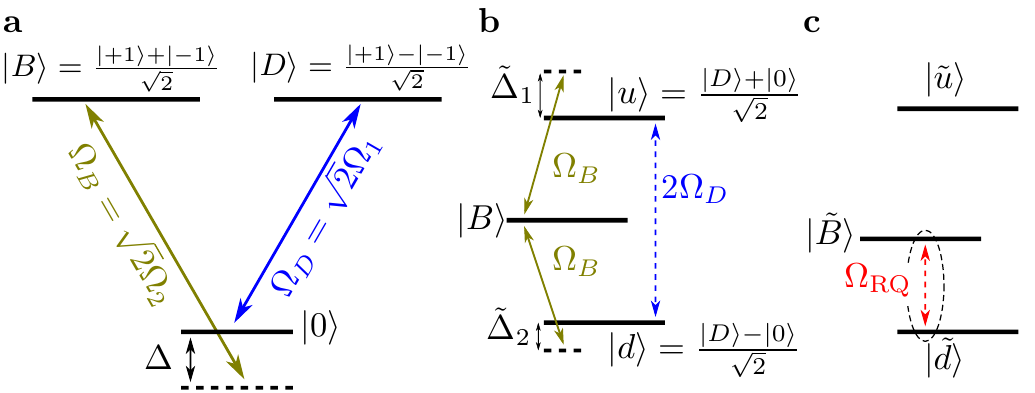}%
    \caption{Bare, dressed and doubly-dressed states in the double lambda drive. 
        Drive fields are indicated by solid lines, whereas energy separations are marked with dashed lines.
        (a) Interaction picture representation of \cref{eq:bare-drive-h} of the bare states in the basis of the drive, $\Omega_1 = \Omega_2 = \Omega$.
        A simultaneous bright ($\ket{B}$) and dark ($\ket{D}$) state drive can be realized by a phase shift of $-\pi$ in the on-resonant lambda scheme in \cref{fig:scheme}a.
        (b) Dressed state picture, where the on-resonant drive is incorporated into the level description.
        The detuning transforms in this picture to $\tilde{\Delta}_1= \Delta + \Omega_1$ and $\tilde{\Delta}_2 = \Delta - \Omega_1$.
        (c) The final doubly-dressed states with incorporated off-resonant drives, where the robust qubit energy gap, $\Omega_{\text{RQ}}=\Omega_{\ket{\tilde{B}},\ket{\tilde{d}}}$, becomes apparent.
    }
    \label{fig:scheme-dressed}
\end{figure}

We consider a three-level system with states $\ket{0}$ and $\ket{\pm1}$, where the $\ket{\pm1}$ states are dipole coupled to the $\ket{0}$ state, as illustrated in \cref{fig:scheme}a. 
The energy gaps ($\hbar=1$) between the $\ket{0}$ state and the $\ket{\pm1}$ states are $ \omega_1$ and $\omega_2$, respectively.
Our scheme utilises four driving fields which can be expressed by the driving Hamiltonian
\begin{equation}
\begin{split}
    H &=  2 \Omega_1 
   \Bigl( \cos( \omega_1 t) \ketbra{0}{-1}
    - \cos( \omega_2 t) \ketbra{0}{1}
    + \text{h.c.} \Bigr)\\
      &\hspace{0.3cm}+   2\Omega_2
      \Bigl( \cos( ( \omega_1 + \Delta) t) \ketbra{0}{-1} \\
& \hspace{1.4cm}    +  \cos( \omega_2 + \Delta) t) \ketbra{0}{1} + \text{h.c.} \Bigr) .
    \label{eq:bare-drive-h}
\end{split}
\end{equation}
Moving to the interaction picture (IP) with respect to 
$H_{01} = \omega_1 \ketbra{-1} + \omega_2 \ketbra{1} $, changing the basis to $\{\ket{0}, \ket{B}, \ket{D}\}$, with $\ket{B} = (\ket{1} + \ket{-1})/\sqrt{2} $ and $\ket{D} = (\ket{1} - \ket{-1})/\sqrt{2} $, and applying the rotating-wave approximation, we obtain the Hamiltonian
\begin{equation}
\begin{split}
        H_I  &=  \sqrt{2}\Omega_1 \bigl( \ketbra{0}{D} +\ketbra{D}{0} \bigr) \\
         &\hspace{0.3cm}+  \sqrt{2}\Omega_2 \bigl( \ketbra{0}{B} e^{i \Delta t} + \ketbra{B}{0} e^{-i \Delta t} \bigr),
\end{split}
    \label{eq:int_state_h}
\end{equation}
which is illustrated by a level scheme depicted in \cref{fig:scheme-dressed}a.
The states $\ket{0}$ and $\ket{D}$ are on-resonantly coupled by a single lambda drive with a strength  $\Omega_D=\sqrt{2} \Omega_1$. 
Under the assumption of $\Delta > \Omega_2$, an off-resonant coupling between  $\ket{0}$ and $\ket{B}$ by $\Omega_B=\sqrt{2} \Omega_2$ is obtained.

The drive, $\Omega_D$, in \cref{fig:scheme-dressed}a transforms into the dressed states $\{\ket{u}, \ket{B}, \ket{d}\}$, as schematically illustrated in \cref{fig:scheme-dressed}b, with the corresponding eigenvalues $\{+\Omega_D, 0, -\Omega_D\}$, respectively.
Here, we introduced the states $\ket{u} =  (\ket{0} + \ket{D})/\sqrt{2}$ and $\ket{d} = (\ket{0}-\ket{D})/\sqrt{2}$.

For a strong enough drive, $\Omega_1$, robustness to magnetic noise is obtained.  
As the magnetic noise, $\delta B$, couples between the $\ket{B}$ state and the $\ket{u}$ and $\ket{d}$ states, we can set $\Omega_D$ such that the power spectrum of the noise, $S_{BB}(\Omega_D)$, is much smaller than $1/T_{1}$, where $T_{1}$ is the system lifetime.
This condition ensures that the first order effect of the magnetic noise is negligible.  

The coherence time of the dressed states is then mainly limited by driving amplitude fluctuations, $\Omega \rightarrow \Omega (1 + \delta(t))$, where $\delta(t)$ represents a random noise contribution. 
To additionally obtain robustness to drive-field fluctuations, we consider the effect of the second detuned drive, $\Omega_B$, on the dressed states.
We therefore move to the basis of the dressed states and to the IP with respect to $H_{02}= \Delta\ketbra{B}$, and obtain
\begin{equation} 
\begin{split}
    H_{II}  &=  \Omega_D  \Bigl( \ketbra{u}{u} - \ketbra{d}{d} \Bigr) - \Delta \ketbra{B}\\
&\hspace{0.3cm}+ \frac{\Omega_B}{\sqrt{2}} \Bigl( \bigl( \ketbra{B}{u} + \ketbra{B}{d} \bigr) + \text{h.c.} \Bigr).
\end{split}
\label{eq:doubly_h}
\end{equation}
The eigenstates of $H_{II}$, denoted by $\{ \ket{\tilde{u}}, \ket{\tilde{B}}, \ket{\tilde{d}}\}$, are termed as the doubly-dressed states and their relative level scheme is illustrated in \cref{fig:scheme-dressed}c.

The effect of drive fluctuations can be introduced in $H_{II}$ by replacing $\Omega_i$ with $\Omega_i \, (1+\delta_i)$. 
As both driving fields originate from the same source, we can assume that the noise is mostly correlated. 
Thus, for each set of eigenstates, $\ket{k}$ and $\ket{j}$, we define the driving coherence time as $T_{2}^{\Omega_{k,j}} = \sqrt{2}/(e^{k}_{\delta}-e^{j}_{\delta})$, where $e^{k}_{\delta}$ is the first order term in $\delta_1$ and $\delta_2$ of the eigenvalue expansion of $\ket{k}$. 
For a given driving noise configuration, i.e., for a given relation between $\delta_1$ and $\delta_2$, the driving parameters $\Omega_1$, $\Omega_2$, and $\Delta$ can be chosen such that the driving coherence time of the two negative eigenvalues of $H_{II}$ is $T_{2}^{\Omega_{\ket{\tilde{B}},\ket{\tilde{d}}}} \gg T_1$.
In this case $e^{\tilde{B}}_{\delta}\approx e^{\tilde{d}}_{\delta}$, which means that the transition frequency of the robust qubit is insensitive to first-order driving fluctuations.  
Hence, with moderate driving fields, the coherence time of this doubly-dressed qubit is mainly limited by the second-order effect of the magnetic noise $\sim \delta B^2/\Omega_D$. 
Increasing the strength of the driving fields reduces the second-order effect of the magnetic noise but introduces an increased second-order effect of the drive noise, $\sim \delta \Omega_D^2$.

\begin{figure}[t!]
    \centering
    \includegraphics[width=1.0\columnwidth]{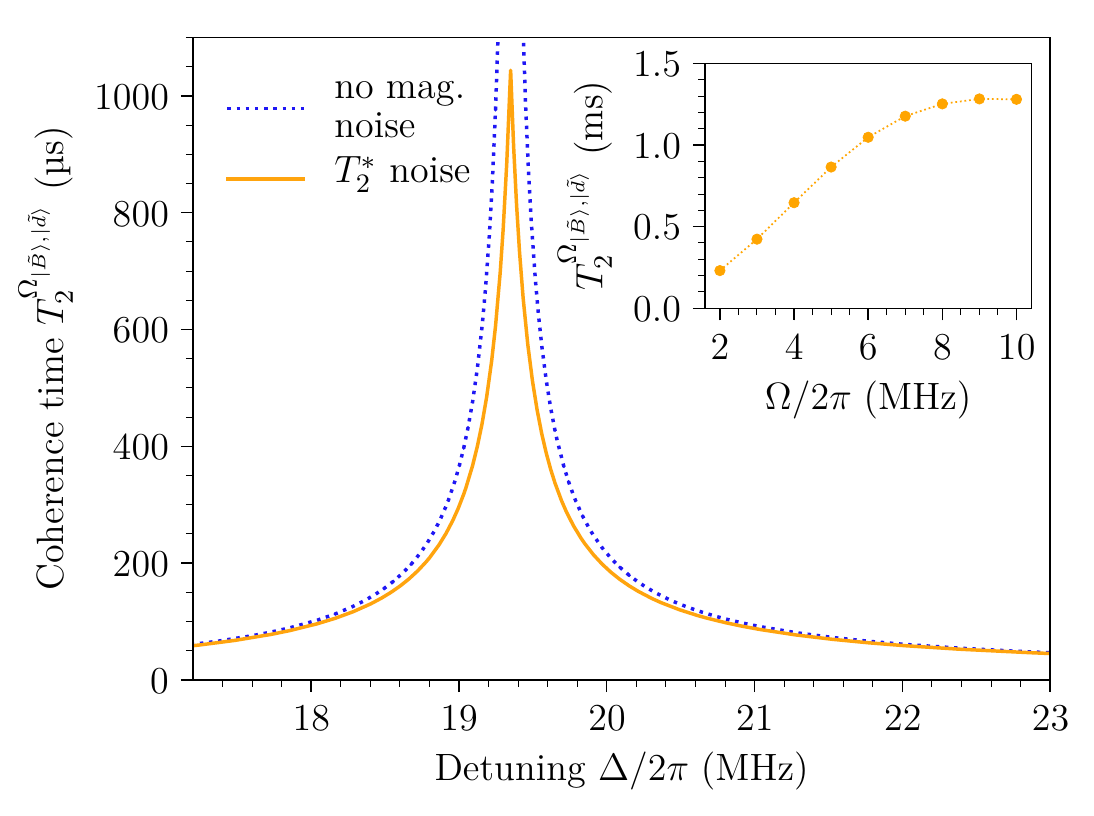}
    \caption{Expected coherence times for a drive  $\Omega_1 = \Omega_2 = 2 \pi \cdot 6$\,MHz and $\delta_1 = \delta_2=0.005$.
        Under the assumption of no magnetic noise, the blue dotted curve predicts the position of optimal detuning, $\Delta/2\pi \approx 19.35$\,MHz, where first-order drive noise of the system is eliminated.
        By including magnetic noise (solid orange line) with a pure dephasing time of $T_2^*=2$\,\textmu s, the coherence time reaches a limit, which is set by the second-order effects of the magnetic and drive noise, and constitutes the maximal improvement of the scheme ($T_{2}^{\Omega_{\ket{\tilde{B}},\ket{\tilde{d}}}} \approx 1$\,ms). 
        Since the coherence time as function of $\Delta$ has a Lorentzian shape, the FWHM of the peak is $\sim 1/ T_{2}^{\Omega_{\ket{\tilde{B}},\ket{\tilde{d}}}}$.
        Inset: Plot of the maximal coherence time obtained for different drive fields, $\Omega = \Omega_1 = \Omega_2$. 
        For increased drive strengths the second-order magnetic noise is reduced but the second-order drive noise is increased. 
        Hence, the coherence time is improved up to $T_{2}^{\Omega_{\ket{\tilde{B}},\ket{\tilde{d}}}} \approx 1.2$\,ms for $\Omega_1=\Omega_2\approx2\pi\cdot 10$\,MHz.  
    }
    \label{fig:theory}
\end{figure}

We consider a NV centre electron spin with a pure dephasing time of $T_2^*\approx 2$\,\textmu s resulting from magnetic noise described by an Ornstein-Uhlenbeck random process \cite{cywinski_how_2008, wang_spin_2013} with a correlation time of $\tau_c \approx 15$\,\textmu s. 
In this case with $\Omega_1=\Omega_2=2\pi\cdot 6$\,MHz a coherence time of $\approx 1$\,ms is obtained (see \cref{fig:theory}), which is limited by the second-order effect of both magnetic and drive noise. 
Further increase of the drive strength, $\Omega$, would allow for a further improvement in the coherence time up to the point where the second-order drive noise is too strong and begins to dominate, which in our case is at  $\Omega_1=\Omega_2\approx2\pi\cdot 10$\,MHz  (see inset of \cref{fig:theory}). 

\section{Experimental results}

The experimental implementation of the proposed scheme follows the protocol illustrated in \cref{fig:scheme}b.
Four drive fields are applied on the bare basis states $\{ \ket{-1},\ket{0}, \ket{1}\}$ of the NV centre (cf. \cref{fig:scheme}a).
The field amplitudes of the on-resonant drives are adjusted to yield an identical Rabi frequency, $\Omega$, for both transitions, $\omega_1$ and $\omega_2$ (see Suppl.~B).
The off-resonant drives are obtained by adding a detuning, $\Delta$, to the resonant drives, resulting in the total field
\begin{equation}
\begin{split}
    \Omega_{\mathrm{tot}}(t) 
&= \Omega \cdot \Big(
\cos ( \omega_1 t )
-  \cos (\omega_2 t )
+ \cos \big( ( \omega_1 +\Delta) t  \big)\\
&\hspace{1.1cm}+  \cos \big( ( \omega_2 +\Delta) t  \big)
\Big) \, ,
\end{split}
\label{eq:drive_fields}
\end{equation}
synthesised by a signal generator ($\Omega_1 = \Omega_2=\Omega = 2 \pi \cdot 2.27$\,MHz) and thereby implementing \cref{eq:bare-drive-h}.
\begin{figure}[!b]
    \centering
    \includegraphics[width=1.0\columnwidth]{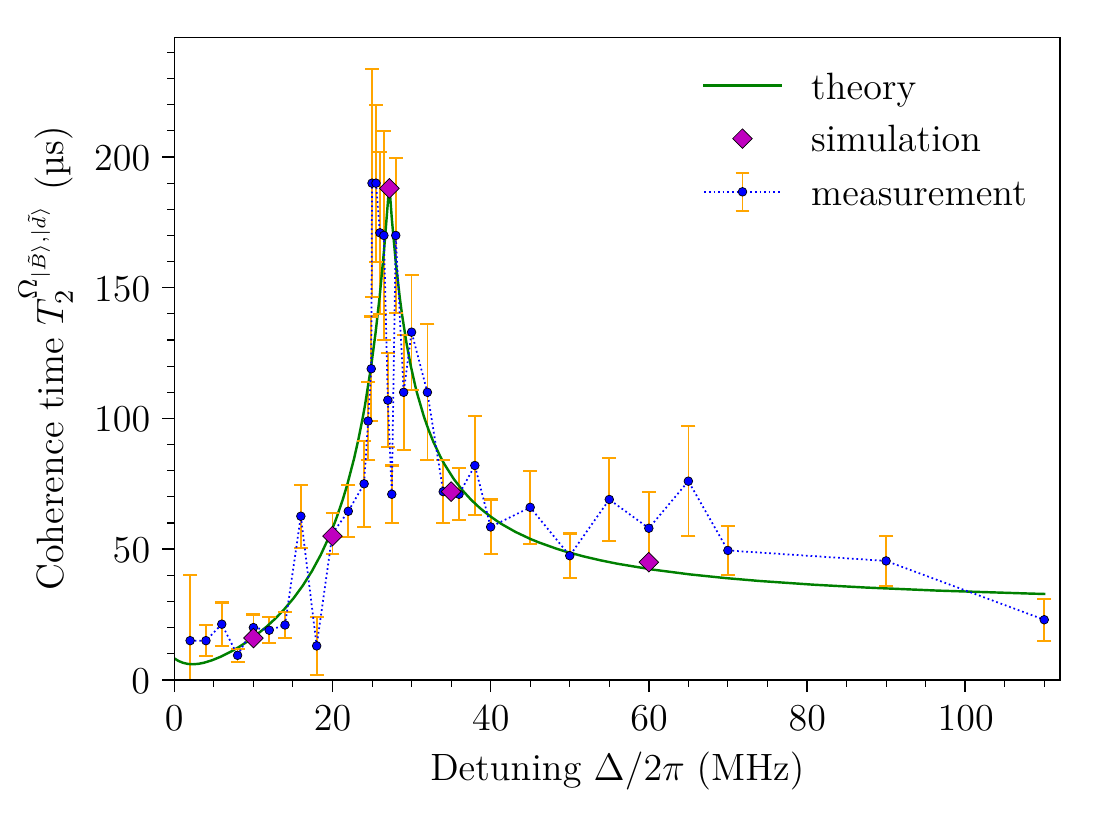}
    \caption{Coherence time measurement, the theoretical model and the simulation of the double lambda drive.
        The record of the coherence time relies on a Ramsey type measurement, performed in the \{$\ket{0},\ket{B},\ket{D}$\}-basis, with resonant and off-resonant drives  corresponding to $\Omega_1=\Omega_2= 2 \pi \cdot 2.27$\,MHz (cf. \cref{fig:scheme}b).
        The theory reproduces the measurement results by considering four times greater drive noise on $\ket{B}$ compared to $\ket{D}$ (cf. \cref{fig:scheme-dressed}a).
        Full simulations were performed for selected points in the graph to verify both, the theoretical model and the measurement.
        The error bars in the measurement curve represent $\Delta T_{2}^{\Omega_{ \ket{\tilde{B}}, \ket{\tilde{d}} }}$ and originate from fitting the coherence time to $\propto\sin(\omega t +\phi)\exp\Bigl\{- \Bigl(t/T_{2}^{\Omega_{ \ket{\tilde{B}}, \ket{\tilde{d}} }} \Bigr) \Bigr\}$.
    }
    \label{fig:main_meas}
\end{figure}
The interaction with the field couples the drive, $\Omega$, to the bare spin states (cf. \cref{fig:scheme}a).
On-resonant drives induce Rabi oscillations at a rate $\Omega_D$, and a
positive (negative) detuning results in 'red' ('blue') detuned AC-Stark shifted energy levels of the $\ket{d}$ and $\ket{B}$ ($\ket{u}$ and $\ket{B}$) states, as shown in \cref{fig:scheme-dressed}b. 
In combination, these drives create the doubly-dressed states, which are depicted in \cref{fig:scheme-dressed}c.
The appearing energy levels of the doubly-dressed states, $\ket{\tilde{u}}, \ket{\tilde{d}}$, and $\ket{\tilde{B}}$, are eventually all coupled to the drive fields (see \cref{fig:scheme-dressed}).

By adjusting the detuning, $\Delta$, a configuration can be obtained, in which two states (either $\ket{\tilde{d}} \leftrightarrow \ket{\tilde{B}}$ or $\ket{\tilde{u}} \leftrightarrow \ket{\tilde{B}}$) experience the same drive noise, $\delta \Omega$.
This eliminates the energy gap fluctuations (due to $\delta \Omega$) between the two considered states and reflects the robustness of the qubit against drive strength fluctuations, $\delta \Omega$.
The large energy gap in the dressed states, $\Omega_D$, which originates from the on-resonant drives, ensures a sufficient decoupling from external magnetic noise contributions, $\delta B$, as it also increases the energy gap of the robust qubit, $\Omega_{\text{RQ}}$ (cf. \cref{fig:scheme-dressed}c).

To determine the performance of the scheme the detuning-dependent coherence times of the protected states have to be recorded, yielding the optimal AC-Stark shifted energy levels that are least sensitive to drive fluctuations.
The measurement is performed in the dressed-state basis and is analogous to a free induction decay (FID) or Ramsey measurement (cf. \cref{fig:scheme}b).
Here, by the application of an on-resonant $\pi/2$ pulse with $\Omega_D$, a superposition is created between the $\ket{B}$ and $\ket{0}$ states (which is also a superposition of the $\ket{D}$ and $\ket{0}$ states, but with a different initial phase factor in the $\ket{-1}$ state).
In the next step, the double lambda drive (depicted in \cref{fig:scheme}a) is applied on the superposition states as a function of interaction time $\tau$, revealing the present energy gaps (in \cref{fig:scheme-dressed}c) as a coherent evolution.
By mapping the coherences to populations with a consecutive $\pi/2$ pulse (with $\Omega_D$), spin-state dependent fluorescence is observed upon the application of a laser pulse.
Finally, the recorded readout signal, $S(\tau)$, contains frequency components proportional to the energy gaps of the doubly-dressed states (between the states in \cref{fig:scheme-dressed}c).
The robust state is identified by the longest measured coherence time in \cref{fig:main_meas} as energy fluctuations in the robust state are significantly suppressed while the oscillations induced by the other energy gaps decay quickly.

The measured coherence times, $T_2^{\Omega_{ \ket{\tilde{B}}, \ket{\tilde{d}}  }}$, are extracted by fitting a sinusoidal exponential decay to $S(\tau)$ and plotted as a function of detuning, $\Delta$, in \cref{fig:main_meas}.
The asymmetric shape of the curve provides insights about the appearing dynamics.
Starting from the limit of a very large detuning, $\Delta$, only the dressed states $\ket{u}$, $\ket{d}$ and $\ket{B}$ are present in the scheme and the AC-Stark shifts are negligible (see \cref{fig:scheme-dressed}b).
Thus, the coherence time of the robust qubit is then predominantly determined by the noise, $\delta \Omega$, as $\delta B$ has a smaller impact at this drive field strength, $\Omega$ (cf. Suppl. S2.B).
Decreasing the detuning, $\Delta$, introduces Stark shifts on the $\ket{u}, \ket{d}$ and $\ket{B}$ states, which effectively reduces the fluctuations of the doubly-dressed state energy gap, thereby prolonging the coherence time of the robust qubit.
Hence, the height of the peak is now mainly limited by second-order noise contributions from $\delta B$ (see the obtained peak in \cref{fig:main_meas}).
Further decrease of the detuning starts to drive $\ket{0} \leftrightarrow \ket{B}$ (cf. \cref{fig:scheme-dressed}a) and introduces thereby again more drive noise, which impacts strongly on the doubly-dressed states.
Approaching zero detuning will eventually eliminate the doubly-dressed states, $\ket{\tilde{u}}$, $\ket{\tilde{d}}$ and $\ket{\tilde{B}}$, and at $\Delta=0$ only the two-level system transition, $\omega_1$, remains to be addressed by $2\Omega$ (cf. \cref{eq:drive_fields} and \cref{fig:scheme}a).
The same behaviour as described above is expected for a negative detuning, $\Delta$.

The theoretical model describing the coherence time as a function of detuning allows us to introduce three parameters in order to mimic the experimental situation.
These are the drive noise, $\delta \Omega_B$ and $\delta \Omega_D$, on the dressed states (cf. \cref{fig:scheme}b) and the magnetic noise, $\delta B$.
It is important to note that these noise parameters do not change the asymmetric shape of the curve as the shape is fully determined by the model.
However, an imbalance between the drive noise strengths impacts the optimal detuning, whereas the magnetic noise solely impacts the attainable coherence time.

The measured free induction decay time, $T_2^* = (1.78 \pm 24)$\,\textmu s, of the bare states sets the theoretical limited for the coherence time of $\sim190$\,\textmu s when the drive noise is eliminated.
By selecting a four times higher noise, $\delta \Omega_B \approx 4\, \delta \Omega_D$, the theoretical dependence of the coherence time on the detuning plotted in \cref{fig:main_meas} is obtained. 
The analytical function of the coherence time is given by $T_{2}^{\Omega_{ \ket{\tilde{B}}, \ket{\tilde{d}} }} = \sqrt{2}/\Bigl((\sqrt{2}/190)+\abs{e^{\tilde{B}}_{\delta}-e^{\tilde{d}}_{\delta}}\Bigr)$ \,\textmu s.
In addition, simulations with magnetic and driving noise models were preformed for several detuning values and that reproduce the experimental results very well. These simulation results are presented in \cref{fig:main_meas} (for more details on the simulations see supplementary). 
In the following, we clarify how quantitatively the noise parameters are grasped in the experiment.

It appears that the DC and AC components of the magnetic noise, $\delta B$, have equal contributions to both transitions, $\omega_1$ and $\omega_2$, as we obtain (within the error bar) the same values for $T_2^*$ and $T_2 = (215 \pm 31)$\,\textmu s (see Suppl. S2.C and S2.D).
However, by comparing the coherence time of the Rabi drives, we obtain a drastic difference, $T_2^{\Omega(\omega_1)}= (62 \pm 12)$\,\textmu s and $T_2^{\Omega(\omega_2)}= (159 \pm 24)$\,\textmu s, which hints at a drive frequency dependent noise spectrum.
As all the fields are produced by the same signal generator, it is valid to consider correlations in the drive noise.
The combination of both effects can truly cause the noise imbalance between $\delta \Omega_B$ and $\delta \Omega_D$, which directly affects the position of the peak with respect to the detuning, $\Delta$.
As this is a setup specific setting, the AC-Stark shifts have to be adjusted to compensate for this value.
However, the coherence time improving effect, as theoretically predicted, is expected to be within the range of 50\,MHz at the utilised drive field, $\Omega$, as larger detunings have a negligible energy shift on the states.
A further and more detailed investigation of the drive noise is presented in Suppl. S3.

\section*{Improved scheme}

So far, our scheme shows how to eliminate the first-order effect of the drive fluctuations, $\delta \Omega$, where for moderate drive fields the coherence time is mainly limited by the second-order effect of the magnetic noise $\sim \delta B^2/\Omega$. 
However, the second-order effect of the magnetic noise can be suppressed in a similar way as demonstrated for the elimination of the first-order drive fluctuations, $\delta \Omega$.

To see this, we consider the on-resonant drive ($\Omega_1$ in \cref{fig:scheme-dressed}). 
For the dressed states, the second-order effect of the magnetic noise is given by $\sim \delta B^2/\Omega_1(\ketbra{u}-\ketbra{d})$, which describes the fluctuation of the robust energy gap (between $\ket{B}$ and $\ket{d}$) with $\sim \delta B^2/\Omega_1$. 
By introducing a one-photon detuning, $\Delta_0$, which denotes a detuning of the coupling between $\ket{D}$ and $\ket{0}$, the symmetry is broken. 
In this case the second-order effect of the magnetic noise is given by $\sim \delta B^2/\Omega_1 (a \ketbra{u} - b \ketbra{d} - c\ketbra{B})$. 
By adjusting the one-photon detuning, $\Delta_0$, we can set $b=c$, and achieve a clock transition that is insensitive to magnetic field fluctuations, $\delta B$ (up to second-order). 

We now combine this idea with the presented elimination of the first-order drive fluctuations, $\delta \Omega$, to obtain a true clock transition. 
Including the one-photon detuning, $\Delta_0$, in the driving fields of both $\Lambda$ systems and magnetic noise, which is given by $\delta B S_z=\delta B (\ketbra{+1}-\ketbra{-1})$, \cref{eq:doubly_h} results in
\begin{equation} 
\begin{split}
    H_{II}'  &=  \Omega_D  \Bigl( \ketbra{u}{u} - \ketbra{d}{d} \Bigr) - \Delta \ketbra{B} - \Delta_0 \ketbra{0}\\
&\hspace{0.3cm}+ \frac{\Omega_B}{\sqrt{2}} \Bigl( \bigl( \ketbra{B}{u} + \ketbra{B}{d} \bigr) + \text{h.c.} \Bigr)\\
&\hspace{0.3cm}+ \frac{\delta B}{\sqrt{2}} \Bigl( e^{i\Delta t}\bigl( \ketbra{B}{u} + \ketbra{B}{d} \bigr) + \text{h.c.} \Bigr),
\end{split}
\label{eq:doubly_h2}
\end{equation}
\begin{figure}[!t]
    \centering
    \includegraphics[width=1.0\columnwidth]{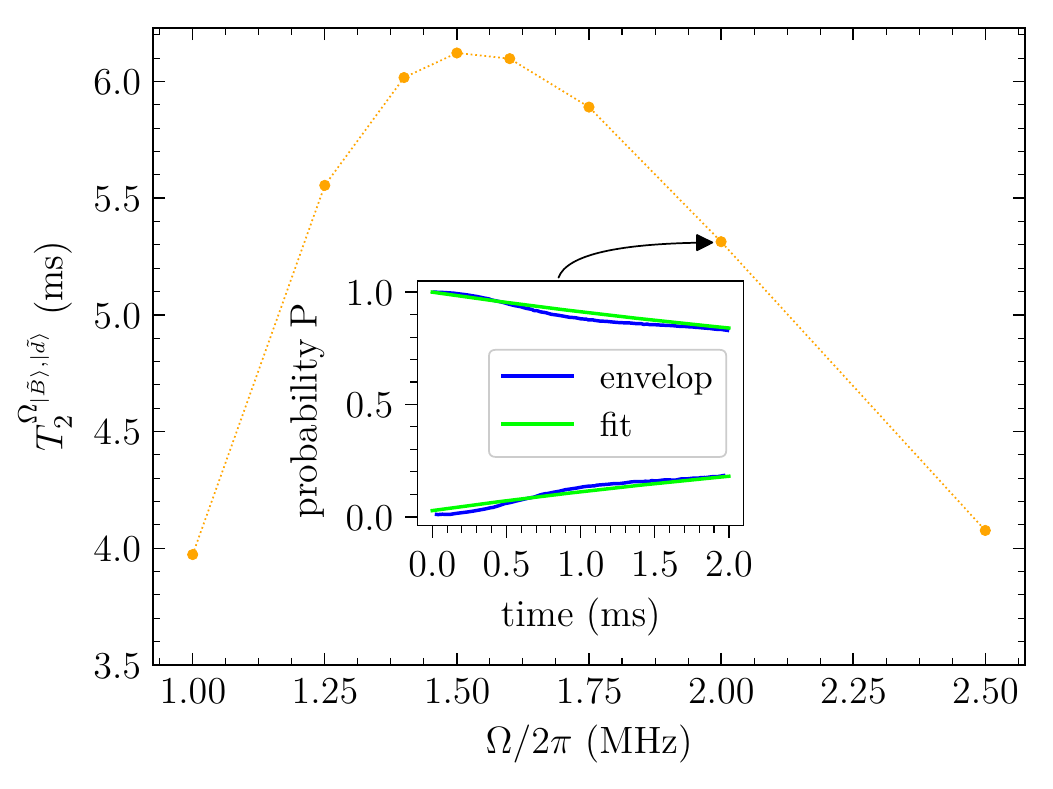}
    \caption{Plot of the estimated maximal coherence time for different drive fields, $\Omega$, assuming   $\delta_1 = \delta_2=0.005$ and the same magnetic noise model as considered for the simulations of the experimental results with $T_2^* \approx 2$\,\textmu s. 
    For increased drive strengths the fourth-order magnetic noise is reduced but the drive noise contributions (second-order terms and amplitude mixing) are increased.
    Inset: Simulation results. Envelope function of the initial state $\ket{\psi_i} = (\ket{\tilde{B}} + \ket{\tilde{d}})/\sqrt{2}$ probability (blue lines), averaged over 64 trials with $\Omega_1 = \Omega_2 = 2 \pi \cdot 2$\,MHz, $\Delta=2 \pi \cdot 8.996$\,MHz, and $\Delta_0=2 \pi \cdot 1.739$\,MHz. 
    For comparison, an exponentially decaying (rising) curve with a time constant of 5.3ms is plotted in green.
    }
    \label{fig:improved_scheme}
\end{figure}
and the IP is now obtained with respect to $H_{02}'= \Delta \ketbra{B} + \Delta_0 \ketbra{0}$.
We continue by moving to the basis of the eigenstates of the drives (the double dressed states)
\begin{equation} 
\begin{split}
    H_{II}''  &\approx  
E_{\tilde{u}} \ketbra{\tilde{u}}{\tilde{u}} 
+ E_{\tilde{B}} \ketbra{\tilde{B}}{\tilde{B}} 
+ E_{\tilde{d}} \ketbra{\tilde{d}}{\tilde{d}} \\
&\hspace{0.3cm}+ \frac{\delta B}{\sqrt{2}} \Bigl( e^{i\Delta t}\bigl(\alpha \ketbra{\tilde{B}}{\tilde{u}} 
+ \beta \ketbra{\tilde{B}}{\tilde{d}} \bigr) + \text{h.c.} \Bigr),
\end{split}
\label{eq:doubly_h2_2}
\end{equation}
where $E_i$ are the eigenvalues, and $\alpha$ and $\beta$ are real coefficients. 
The drive noise is treated as before, where we require $e^{\tilde{B}}_{\delta}\approx e^{\tilde{d}}_{\delta}$. 
This gives us one constraint on $\Delta$ and $\Delta_0$. 
The second constraint comes from the elimination of the second-order effect of the magnetic noise.
Moving to the IP with respect to $H_{03}'= -\Delta \ketbra{\tilde{B}}$ we get the time independent Hamiltonian  
\begin{equation} 
\begin{split}
    H_{III}''  &\approx  E_{\tilde{u}} \ketbra{\tilde{u}}{\tilde{u}} 
+ \bigl(E_{\tilde{B}} + \Delta\bigr) \ketbra{\tilde{B}}{\tilde{B}} 
+ E_{\tilde{d}}\ketbra{\tilde{d}}{\tilde{d}} \\
&\hspace{0.3cm}+ \frac{\delta B}{\sqrt{2}} \Bigl(\bigl(\alpha \ketbra{\tilde{B}}{\tilde{u}} 
+ \beta \ketbra{\tilde{B}}{\tilde{d}} \bigr) + \text{h.c.} \Bigr),
\end{split}
\label{eq:doubly_h2_3}
\end{equation}
This enables the calculation of the second-order contribution of the magnetic noise to the eigenvalues, which are the $b$ and $c$ coefficients, as a function of $\Delta$ and $\Delta_0$. 
The two constraints, $e^{\tilde{B}}_{\delta}\approx e^{\tilde{d}}_{\delta}$ and  $b\approx c$, allow us to determine the optimal values of $\Delta$ and $\Delta_0$. 
For the considered drive noise, $\delta_1 = \delta_2=0.005$, in \cref{fig:theory}, and for a moderate drive of $\Omega_1 = \Omega_2 = 2 \pi \cdot 2$\,MHz, the optimal detunings would be obtained by $\Delta=2 \pi \cdot8.9956$\,MHz and $\Delta_0=2 \pi \cdot1.7386$\,MHz. 
We simulated the present driving configuration under the effect of the same magnetic noise model considered for the simulations of the experiment in \cref{fig:main_meas} (with $T_2^* \approx 2$\,\textmu s). 
The results of the simulation are shown in the inset of \cref{fig:improved_scheme} indicating an improvement of more than $1$ order of magnitude in the coherence time compared to the original scheme. 
There are two limiting factors on the coherence time. 
The first factor are the higher-order terms of the noise; the second-order term of the driving noise and the fourth order term of the magnetic noise. 
With the parameters considered in the simulation these terms result in a limit of $\sim 10$ ms on the coherence time. 
The second factor is the amplitude mixing between the eigenstates due to fast rotating terms, which introduces first order driving noise. 
In our case the mixing is $\sim 0.2\%$, which means that the coherence time is limited to 500 times the driving noise limited coherence time($\sim 20$ $\mu$s here), and hence the limit on the coherence time is  $\sim 10$ ms. Therefore, taking both factors into account we conclude that the coherence time is limited to $\sim 5.3$ ms, which is in agreement with the simulation results. 
For low drive fields, the drive noise contributions (second-order terms and amplitude mixing) is small and the coherence time is mainly limited by the fourth-order magnetic noise. 
For higher drive fields the fourth order magnetic noise becomes negligible and the coherence time is mainly limited by the drive noise contributions. 
The estimated maximal coherence time for different drive fields strength is shown in \cref{fig:improved_scheme}. 
The amplitude mixing can also be decreased by increasing the Zeeman splitting. Given the noise parameters, one can optimise the driving parameters with respect to these factors and obtain the optimal coherence time.  

\section*{Conclusion}

In this work we presented and experimentally demonstrated a new scheme for the creation of a robust qubit in a three-level system by means of a clock transition adjustment. 
The basic scheme is based on the application of continuous resonant and off-resonant drive fields. 
The resonant drive fields provide robustness to environmental noise whereas the off-resonant drive fields eliminate the first-order effect of the drive noise, by tuning a clock like transition that is insensitive to first-order shifts of the drive-field amplitudes.
For the case of the NV centre in diamond, we achieved an improvement of $\sim 2$ orders of magnitude in the coherence time compared to the pure dephasing time while utilising moderate drive fields.
In the optimal version of the scheme, the clock transition adjustment is extended to also eliminate the second-order effect of the environmental noise without necessitating strong drive fields. 
Hence, our scheme enables prolonged coherence times reaching the lifetime limit using a relatively simple experimental setup and without requiring extremely strong drive fields.

This scheme facilitates the sensing of AC magnetic fields, and in particular, high frequency fields in the GHz regime, where the sensitivity would be solely limited by the coherence time of the robust qubit.
While this work has focused solely on the NV centre, we believe that this scheme is applicable to a variety of atomic and solid-state systems with optical or microwave transitions, such as trapped ions, rare-earth ions, and other defect centres.
Therefore, we believe that the scheme has potential applications in a wide range of tasks in the fields of quantum information science and technology, and in particular quantum sensing.

\section*{Acknowledgements}

The experiments presented here were realised by the Qudi Software Suite \cite{binder_qudi_2017}.
We would like to thank Christian Osterkamp and Kristian Hagsted Rasmussen for the sample preparation.
A.S., A.H., H.A.R.E.-E. and U.L.A. acknowledge fundings from the Innovation Foundation Denmark through the project EXMAD and the Qubiz center, the Danish National Research Foundation, and the Danish Research Council via the Sapere Aude project (DIMS).
A.R. acknowledges the support of the Israel Science Foundation (grant no. 1500/13)
 
%


\begin{thebibliography}{46}%
	\makeatletter
	\providecommand \@ifxundefined [1]{%
		\@ifx{#1\undefined}
	}%
	\providecommand \@ifnum [1]{%
		\ifnum #1\expandafter \@firstoftwo
		\else \expandafter \@secondoftwo
		\fi
	}%
	\providecommand \@ifx [1]{%
		\ifx #1\expandafter \@firstoftwo
		\else \expandafter \@secondoftwo
		\fi
	}%
	\providecommand \natexlab [1]{#1}%
	\providecommand \enquote  [1]{``#1''}%
	\providecommand \bibnamefont  [1]{#1}%
	\providecommand \bibfnamefont [1]{#1}%
	\providecommand \citenamefont [1]{#1}%
	\providecommand \href@noop [0]{\@secondoftwo}%
	\providecommand \href [0]{\begingroup \@sanitize@url \@href}%
	\providecommand \@href[1]{\@@startlink{#1}\@@href}%
	\providecommand \@@href[1]{\endgroup#1\@@endlink}%
	\providecommand \@sanitize@url [0]{\catcode `\\12\catcode `\$12\catcode
		`\&12\catcode `\#12\catcode `\^12\catcode `\_12\catcode `\%12\relax}%
	\providecommand \@@startlink[1]{}%
	\providecommand \@@endlink[0]{}%
	\providecommand \url  [0]{\begingroup\@sanitize@url \@url }%
	\providecommand \@url [1]{\endgroup\@href {#1}{\urlprefix }}%
	\providecommand \urlprefix  [0]{URL }%
	\providecommand \Eprint [0]{\href }%
	\providecommand \doibase [0]{http://dx.doi.org/}%
	\providecommand \selectlanguage [0]{\@gobble}%
	\providecommand \bibinfo  [0]{\@secondoftwo}%
	\providecommand \bibfield  [0]{\@secondoftwo}%
	\providecommand \translation [1]{[#1]}%
	\providecommand \BibitemOpen [0]{}%
	\providecommand \bibitemStop [0]{}%
	\providecommand \bibitemNoStop [0]{.\EOS\space}%
	\providecommand \EOS [0]{\spacefactor3000\relax}%
	\providecommand \BibitemShut  [1]{\csname bibitem#1\endcsname}%
	\let\auto@bib@innerbib\@empty
	\bibitem [{\citenamefont {Hahn}(1950)}]{hahn_spin_1950}%
	\BibitemOpen
	\bibfield  {author} {\bibinfo {author} {\bibfnamefont {E.~L.}\ \bibnamefont
			{Hahn}},\ }\href {\doibase 10.1103/PhysRev.80.580} {\bibfield  {journal}
		{\bibinfo  {journal} {Physical Review}\ }\textbf {\bibinfo {volume} {80}},\
		\bibinfo {pages} {580} (\bibinfo {year} {1950})}\BibitemShut {NoStop}%
	\bibitem [{\citenamefont {Carr}\ and\ \citenamefont
		{Purcell}(1954)}]{carr_effects_1954}%
	\BibitemOpen
	\bibfield  {author} {\bibinfo {author} {\bibfnamefont {H.~Y.}\ \bibnamefont
			{Carr}}\ and\ \bibinfo {author} {\bibfnamefont {E.~M.}\ \bibnamefont
			{Purcell}},\ }\href {\doibase 10.1103/PhysRev.94.630} {\bibfield  {journal}
		{\bibinfo  {journal} {Physical Review}\ }\textbf {\bibinfo {volume} {94}},\
		\bibinfo {pages} {630} (\bibinfo {year} {1954})}\BibitemShut {NoStop}%
	\bibitem [{\citenamefont {Meiboom}\ and\ \citenamefont
		{Gill}(1958)}]{meiboom_modified_1958}%
	\BibitemOpen
	\bibfield  {author} {\bibinfo {author} {\bibfnamefont {S.}~\bibnamefont
			{Meiboom}}\ and\ \bibinfo {author} {\bibfnamefont {D.}~\bibnamefont {Gill}},\
	}\href {\doibase 10.1063/1.1716296} {\bibfield  {journal} {\bibinfo
			{journal} {Review of Scientific Instruments}\ }\textbf {\bibinfo {volume}
			{29}},\ \bibinfo {pages} {688} (\bibinfo {year} {1958})}\BibitemShut
	{NoStop}%
	\bibitem [{\citenamefont {Viola}\ and\ \citenamefont
		{Lloyd}(1998)}]{viola_dynamical_1998}%
	\BibitemOpen
	\bibfield  {author} {\bibinfo {author} {\bibfnamefont {L.}~\bibnamefont
			{Viola}}\ and\ \bibinfo {author} {\bibfnamefont {S.}~\bibnamefont {Lloyd}},\
	}\href {\doibase 10.1103/PhysRevA.58.2733} {\bibfield  {journal} {\bibinfo
			{journal} {Physical Review A}\ }\textbf {\bibinfo {volume} {58}},\ \bibinfo
		{pages} {2733} (\bibinfo {year} {1998})}\BibitemShut {NoStop}%
	\bibitem [{\citenamefont {Biercuk}\ \emph {et~al.}(2009)\citenamefont
		{Biercuk}, \citenamefont {Uys}, \citenamefont {VanDevender}, \citenamefont
		{Shiga}, \citenamefont {Itano},\ and\ \citenamefont
		{Bollinger}}]{biercuk_optimized_2009}%
	\BibitemOpen
	\bibfield  {author} {\bibinfo {author} {\bibfnamefont {M.~J.}\ \bibnamefont
			{Biercuk}}, \bibinfo {author} {\bibfnamefont {H.}~\bibnamefont {Uys}},
		\bibinfo {author} {\bibfnamefont {A.~P.}\ \bibnamefont {VanDevender}},
		\bibinfo {author} {\bibfnamefont {N.}~\bibnamefont {Shiga}}, \bibinfo
		{author} {\bibfnamefont {W.~M.}\ \bibnamefont {Itano}}, \ and\ \bibinfo
		{author} {\bibfnamefont {J.~J.}\ \bibnamefont {Bollinger}},\ }\href {\doibase
		10.1038/nature07951} {\bibfield  {journal} {\bibinfo  {journal} {Nature}\
		}\textbf {\bibinfo {volume} {458}},\ \bibinfo {pages} {996} (\bibinfo {year}
		{2009})}\BibitemShut {NoStop}%
	\bibitem [{\citenamefont {Du}\ \emph {et~al.}(2009)\citenamefont {Du},
		\citenamefont {Rong}, \citenamefont {Zhao}, \citenamefont {Wang},
		\citenamefont {Yang},\ and\ \citenamefont {Liu}}]{du_preserving_2009}%
	\BibitemOpen
	\bibfield  {author} {\bibinfo {author} {\bibfnamefont {J.}~\bibnamefont
			{Du}}, \bibinfo {author} {\bibfnamefont {X.}~\bibnamefont {Rong}}, \bibinfo
		{author} {\bibfnamefont {N.}~\bibnamefont {Zhao}}, \bibinfo {author}
		{\bibfnamefont {Y.}~\bibnamefont {Wang}}, \bibinfo {author} {\bibfnamefont
			{J.}~\bibnamefont {Yang}}, \ and\ \bibinfo {author} {\bibfnamefont {R.~B.}\
			\bibnamefont {Liu}},\ }\href {\doibase 10.1038/nature08470} {\bibfield
		{journal} {\bibinfo  {journal} {Nature}\ }\textbf {\bibinfo {volume} {461}},\
		\bibinfo {pages} {1265} (\bibinfo {year} {2009})}\BibitemShut {NoStop}%
	\bibitem [{\citenamefont {Lange}\ \emph {et~al.}(2010)\citenamefont {Lange},
		\citenamefont {Wang}, \citenamefont {Ristè}, \citenamefont {Dobrovitski},\
		and\ \citenamefont {Hanson}}]{lange_universal_2010}%
	\BibitemOpen
	\bibfield  {author} {\bibinfo {author} {\bibfnamefont {G.~d.}\ \bibnamefont
			{Lange}}, \bibinfo {author} {\bibfnamefont {Z.~H.}\ \bibnamefont {Wang}},
		\bibinfo {author} {\bibfnamefont {D.}~\bibnamefont {Ristè}}, \bibinfo
		{author} {\bibfnamefont {V.~V.}\ \bibnamefont {Dobrovitski}}, \ and\ \bibinfo
		{author} {\bibfnamefont {R.}~\bibnamefont {Hanson}},\ }\href {\doibase
		10.1126/science.1192739} {\bibfield  {journal} {\bibinfo  {journal}
			{Science}\ }\textbf {\bibinfo {volume} {330}},\ \bibinfo {pages} {60}
		(\bibinfo {year} {2010})}\BibitemShut {NoStop}%
	\bibitem [{\citenamefont {Ryan}\ \emph {et~al.}(2010)\citenamefont {Ryan},
		\citenamefont {Hodges},\ and\ \citenamefont {Cory}}]{ryan_robust_2010}%
	\BibitemOpen
	\bibfield  {author} {\bibinfo {author} {\bibfnamefont {C.~A.}\ \bibnamefont
			{Ryan}}, \bibinfo {author} {\bibfnamefont {J.~S.}\ \bibnamefont {Hodges}}, \
		and\ \bibinfo {author} {\bibfnamefont {D.~G.}\ \bibnamefont {Cory}},\ }\href
	{\doibase 10.1103/PhysRevLett.105.200402} {\bibfield  {journal} {\bibinfo
			{journal} {Physical Review Letters}\ }\textbf {\bibinfo {volume} {105}},\
		\bibinfo {pages} {200402} (\bibinfo {year} {2010})}\BibitemShut {NoStop}%
	\bibitem [{\citenamefont {Naydenov}\ \emph {et~al.}(2011)\citenamefont
		{Naydenov}, \citenamefont {Dolde}, \citenamefont {Hall}, \citenamefont
		{Shin}, \citenamefont {Fedder}, \citenamefont {Hollenberg}, \citenamefont
		{Jelezko},\ and\ \citenamefont {Wrachtrup}}]{naydenov_dynamical_2011}%
	\BibitemOpen
	\bibfield  {author} {\bibinfo {author} {\bibfnamefont {B.}~\bibnamefont
			{Naydenov}}, \bibinfo {author} {\bibfnamefont {F.}~\bibnamefont {Dolde}},
		\bibinfo {author} {\bibfnamefont {L.~T.}\ \bibnamefont {Hall}}, \bibinfo
		{author} {\bibfnamefont {C.}~\bibnamefont {Shin}}, \bibinfo {author}
		{\bibfnamefont {H.}~\bibnamefont {Fedder}}, \bibinfo {author} {\bibfnamefont
			{L.~C.~L.}\ \bibnamefont {Hollenberg}}, \bibinfo {author} {\bibfnamefont
			{F.}~\bibnamefont {Jelezko}}, \ and\ \bibinfo {author} {\bibfnamefont
			{J.}~\bibnamefont {Wrachtrup}},\ }\href {\doibase 10.1103/PhysRevB.83.081201}
	{\bibfield  {journal} {\bibinfo  {journal} {Physical Review B}\ }\textbf
		{\bibinfo {volume} {83}},\ \bibinfo {pages} {081201} (\bibinfo {year}
		{2011})}\BibitemShut {NoStop}%
	\bibitem [{\citenamefont {Wang}\ \emph {et~al.}(2012)\citenamefont {Wang},
		\citenamefont {de~Lange}, \citenamefont {Ristè}, \citenamefont {Hanson},\
		and\ \citenamefont {Dobrovitski}}]{wang_comparison_2012}%
	\BibitemOpen
	\bibfield  {author} {\bibinfo {author} {\bibfnamefont {Z.-H.}\ \bibnamefont
			{Wang}}, \bibinfo {author} {\bibfnamefont {G.}~\bibnamefont {de~Lange}},
		\bibinfo {author} {\bibfnamefont {D.}~\bibnamefont {Ristè}}, \bibinfo
		{author} {\bibfnamefont {R.}~\bibnamefont {Hanson}}, \ and\ \bibinfo {author}
		{\bibfnamefont {V.~V.}\ \bibnamefont {Dobrovitski}},\ }\href {\doibase
		10.1103/PhysRevB.85.155204} {\bibfield  {journal} {\bibinfo  {journal}
			{Physical Review B}\ }\textbf {\bibinfo {volume} {85}},\ \bibinfo {pages}
		{155204} (\bibinfo {year} {2012})}\BibitemShut {NoStop}%
	\bibitem [{\citenamefont {Bar-Gill}\ \emph {et~al.}(2013)\citenamefont
		{Bar-Gill}, \citenamefont {Pham}, \citenamefont {Jarmola}, \citenamefont
		{Budker},\ and\ \citenamefont {Walsworth}}]{bar-gill_solid-state_2013}%
	\BibitemOpen
	\bibfield  {author} {\bibinfo {author} {\bibfnamefont {N.}~\bibnamefont
			{Bar-Gill}}, \bibinfo {author} {\bibfnamefont {L.~M.}\ \bibnamefont {Pham}},
		\bibinfo {author} {\bibfnamefont {A.}~\bibnamefont {Jarmola}}, \bibinfo
		{author} {\bibfnamefont {D.}~\bibnamefont {Budker}}, \ and\ \bibinfo {author}
		{\bibfnamefont {R.~L.}\ \bibnamefont {Walsworth}},\ }\href {\doibase
		10.1038/ncomms2771} {\bibfield  {journal} {\bibinfo  {journal} {Nature
				Communications}\ }\textbf {\bibinfo {volume} {4}},\ \bibinfo {pages} {1743}
		(\bibinfo {year} {2013})}\BibitemShut {NoStop}%
	\bibitem [{\citenamefont {Baumgart}\ \emph {et~al.}(2016)\citenamefont
		{Baumgart}, \citenamefont {Cai}, \citenamefont {Retzker}, \citenamefont
		{Plenio},\ and\ \citenamefont {Wunderlich}}]{baumgart_ultrasensitive_2016}%
	\BibitemOpen
	\bibfield  {author} {\bibinfo {author} {\bibfnamefont {I.}~\bibnamefont
			{Baumgart}}, \bibinfo {author} {\bibfnamefont {J.-M.}\ \bibnamefont {Cai}},
		\bibinfo {author} {\bibfnamefont {A.}~\bibnamefont {Retzker}}, \bibinfo
		{author} {\bibfnamefont {M.}~\bibnamefont {Plenio}}, \ and\ \bibinfo {author}
		{\bibfnamefont {C.}~\bibnamefont {Wunderlich}},\ }\href {\doibase
		10.1103/PhysRevLett.116.240801} {\bibfield  {journal} {\bibinfo  {journal}
			{Physical Review Letters}\ }\textbf {\bibinfo {volume} {116}},\ \bibinfo
		{pages} {240801} (\bibinfo {year} {2016})}\BibitemShut {NoStop}%
	\bibitem [{\citenamefont {Khodjasteh}\ and\ \citenamefont
		{Lidar}(2005)}]{khodjasteh_fault-tolerant_2005}%
	\BibitemOpen
	\bibfield  {author} {\bibinfo {author} {\bibfnamefont {K.}~\bibnamefont
			{Khodjasteh}}\ and\ \bibinfo {author} {\bibfnamefont {D.~A.}\ \bibnamefont
			{Lidar}},\ }\href {\doibase 10.1103/PhysRevLett.95.180501} {\bibfield
		{journal} {\bibinfo  {journal} {Physical Review Letters}\ }\textbf {\bibinfo
			{volume} {95}},\ \bibinfo {pages} {180501} (\bibinfo {year}
		{2005})}\BibitemShut {NoStop}%
	\bibitem [{\citenamefont {Uhrig}(2007)}]{uhrig_keeping_2007}%
	\BibitemOpen
	\bibfield  {author} {\bibinfo {author} {\bibfnamefont {G.~S.}\ \bibnamefont
			{Uhrig}},\ }\href {\doibase 10.1103/PhysRevLett.98.100504} {\bibfield
		{journal} {\bibinfo  {journal} {Physical Review Letters}\ }\textbf {\bibinfo
			{volume} {98}},\ \bibinfo {pages} {100504} (\bibinfo {year}
		{2007})}\BibitemShut {NoStop}%
	\bibitem [{\citenamefont {Souza}\ \emph {et~al.}(2011)\citenamefont {Souza},
		\citenamefont {Álvarez},\ and\ \citenamefont {Suter}}]{souza_robust_2011}%
	\BibitemOpen
	\bibfield  {author} {\bibinfo {author} {\bibfnamefont {A.~M.}\ \bibnamefont
			{Souza}}, \bibinfo {author} {\bibfnamefont {G.~A.}\ \bibnamefont {Álvarez}},
		\ and\ \bibinfo {author} {\bibfnamefont {D.}~\bibnamefont {Suter}},\ }\href
	{\doibase 10.1103/PhysRevLett.106.240501} {\bibfield  {journal} {\bibinfo
			{journal} {Physical Review Letters}\ }\textbf {\bibinfo {volume} {106}},\
		\bibinfo {pages} {240501} (\bibinfo {year} {2011})}\BibitemShut {NoStop}%
	\bibitem [{\citenamefont {Yang}\ \emph {et~al.}(2011)\citenamefont {Yang},
		\citenamefont {Wang},\ and\ \citenamefont {Liu}}]{yang_preserving_2011}%
	\BibitemOpen
	\bibfield  {author} {\bibinfo {author} {\bibfnamefont {W.}~\bibnamefont
			{Yang}}, \bibinfo {author} {\bibfnamefont {Z.-Y.}\ \bibnamefont {Wang}}, \
		and\ \bibinfo {author} {\bibfnamefont {R.-B.}\ \bibnamefont {Liu}},\ }\href
	{\doibase 10.1007/s11467-010-0113-8} {\bibfield  {journal} {\bibinfo
			{journal} {Frontiers of Physics in China}\ }\textbf {\bibinfo {volume} {6}},\
		\bibinfo {pages} {2} (\bibinfo {year} {2011})}\BibitemShut {NoStop}%
	\bibitem [{\citenamefont {Farfurnik}\ \emph {et~al.}(2015)\citenamefont
		{Farfurnik}, \citenamefont {Jarmola}, \citenamefont {Pham}, \citenamefont
		{Wang}, \citenamefont {Dobrovitski}, \citenamefont {Walsworth}, \citenamefont
		{Budker},\ and\ \citenamefont {Bar-Gill}}]{farfurnik_optimizing_2015}%
	\BibitemOpen
	\bibfield  {author} {\bibinfo {author} {\bibfnamefont {D.}~\bibnamefont
			{Farfurnik}}, \bibinfo {author} {\bibfnamefont {A.}~\bibnamefont {Jarmola}},
		\bibinfo {author} {\bibfnamefont {L.~M.}\ \bibnamefont {Pham}}, \bibinfo
		{author} {\bibfnamefont {Z.~H.}\ \bibnamefont {Wang}}, \bibinfo {author}
		{\bibfnamefont {V.~V.}\ \bibnamefont {Dobrovitski}}, \bibinfo {author}
		{\bibfnamefont {R.~L.}\ \bibnamefont {Walsworth}}, \bibinfo {author}
		{\bibfnamefont {D.}~\bibnamefont {Budker}}, \ and\ \bibinfo {author}
		{\bibfnamefont {N.}~\bibnamefont {Bar-Gill}},\ }\href {\doibase
		10.1103/PhysRevB.92.060301} {\bibfield  {journal} {\bibinfo  {journal}
			{Physical Review B}\ }\textbf {\bibinfo {volume} {92}},\ \bibinfo {pages}
		{060301} (\bibinfo {year} {2015})}\BibitemShut {NoStop}%
	\bibitem [{\citenamefont {Gordon}\ \emph {et~al.}(2008)\citenamefont {Gordon},
		\citenamefont {Kurizki},\ and\ \citenamefont {Lidar}}]{gordon_optimal_2008}%
	\BibitemOpen
	\bibfield  {author} {\bibinfo {author} {\bibfnamefont {G.}~\bibnamefont
			{Gordon}}, \bibinfo {author} {\bibfnamefont {G.}~\bibnamefont {Kurizki}}, \
		and\ \bibinfo {author} {\bibfnamefont {D.~A.}\ \bibnamefont {Lidar}},\ }\href
	{\doibase 10.1103/PhysRevLett.101.010403} {\bibfield  {journal} {\bibinfo
			{journal} {Physical Review Letters}\ }\textbf {\bibinfo {volume} {101}},\
		\bibinfo {pages} {010403} (\bibinfo {year} {2008})}\BibitemShut {NoStop}%
	\bibitem [{\citenamefont {Fanchini}\ \emph {et~al.}(2007)\citenamefont
		{Fanchini}, \citenamefont {Hornos},\ and\ \citenamefont
		{Napolitano}}]{fanchini_continuously_2007}%
	\BibitemOpen
	\bibfield  {author} {\bibinfo {author} {\bibfnamefont {F.~F.}\ \bibnamefont
			{Fanchini}}, \bibinfo {author} {\bibfnamefont {J.~E.~M.}\ \bibnamefont
			{Hornos}}, \ and\ \bibinfo {author} {\bibfnamefont {R.~d.~J.}\ \bibnamefont
			{Napolitano}},\ }\href {\doibase 10.1103/PhysRevA.75.022329} {\bibfield
		{journal} {\bibinfo  {journal} {Physical Review A}\ }\textbf {\bibinfo
			{volume} {75}},\ \bibinfo {pages} {022329} (\bibinfo {year}
		{2007})}\BibitemShut {NoStop}%
	\bibitem [{\citenamefont {Rabl}\ \emph {et~al.}(2009)\citenamefont {Rabl},
		\citenamefont {Cappellaro}, \citenamefont {Dutt}, \citenamefont {Jiang},
		\citenamefont {Maze},\ and\ \citenamefont {Lukin}}]{rabl_strong_2009}%
	\BibitemOpen
	\bibfield  {author} {\bibinfo {author} {\bibfnamefont {P.}~\bibnamefont
			{Rabl}}, \bibinfo {author} {\bibfnamefont {P.}~\bibnamefont {Cappellaro}},
		\bibinfo {author} {\bibfnamefont {M.~V.~G.}\ \bibnamefont {Dutt}}, \bibinfo
		{author} {\bibfnamefont {L.}~\bibnamefont {Jiang}}, \bibinfo {author}
		{\bibfnamefont {J.~R.}\ \bibnamefont {Maze}}, \ and\ \bibinfo {author}
		{\bibfnamefont {M.~D.}\ \bibnamefont {Lukin}},\ }\href {\doibase
		10.1103/PhysRevB.79.041302} {\bibfield  {journal} {\bibinfo  {journal}
			{Physical Review B}\ }\textbf {\bibinfo {volume} {79}},\ \bibinfo {pages}
		{041302} (\bibinfo {year} {2009})}\BibitemShut {NoStop}%
	\bibitem [{\citenamefont {Clausen}\ \emph {et~al.}(2010)\citenamefont
		{Clausen}, \citenamefont {Bensky},\ and\ \citenamefont
		{Kurizki}}]{clausen_bath-optimized_2010}%
	\BibitemOpen
	\bibfield  {author} {\bibinfo {author} {\bibfnamefont {J.}~\bibnamefont
			{Clausen}}, \bibinfo {author} {\bibfnamefont {G.}~\bibnamefont {Bensky}}, \
		and\ \bibinfo {author} {\bibfnamefont {G.}~\bibnamefont {Kurizki}},\ }\href
	{\doibase 10.1103/PhysRevLett.104.040401} {\bibfield  {journal} {\bibinfo
			{journal} {Physical Review Letters}\ }\textbf {\bibinfo {volume} {104}},\
		\bibinfo {pages} {040401} (\bibinfo {year} {2010})}\BibitemShut {NoStop}%
	\bibitem [{\citenamefont {Bermudez}\ \emph {et~al.}(2011)\citenamefont
		{Bermudez}, \citenamefont {Jelezko}, \citenamefont {Plenio},\ and\
		\citenamefont {Retzker}}]{bermudez_electron-mediated_2011}%
	\BibitemOpen
	\bibfield  {author} {\bibinfo {author} {\bibfnamefont {A.}~\bibnamefont
			{Bermudez}}, \bibinfo {author} {\bibfnamefont {F.}~\bibnamefont {Jelezko}},
		\bibinfo {author} {\bibfnamefont {M.~B.}\ \bibnamefont {Plenio}}, \ and\
		\bibinfo {author} {\bibfnamefont {A.}~\bibnamefont {Retzker}},\ }\href
	{\doibase 10.1103/PhysRevLett.107.150503} {\bibfield  {journal} {\bibinfo
			{journal} {Physical Review Letters}\ }\textbf {\bibinfo {volume} {107}},\
		\bibinfo {pages} {150503} (\bibinfo {year} {2011})}\BibitemShut {NoStop}%
	\bibitem [{\citenamefont {Bermudez}\ \emph {et~al.}(2012)\citenamefont
		{Bermudez}, \citenamefont {Schmidt}, \citenamefont {Plenio},\ and\
		\citenamefont {Retzker}}]{bermudez_robust_2012}%
	\BibitemOpen
	\bibfield  {author} {\bibinfo {author} {\bibfnamefont {A.}~\bibnamefont
			{Bermudez}}, \bibinfo {author} {\bibfnamefont {P.~O.}\ \bibnamefont
			{Schmidt}}, \bibinfo {author} {\bibfnamefont {M.~B.}\ \bibnamefont {Plenio}},
		\ and\ \bibinfo {author} {\bibfnamefont {A.}~\bibnamefont {Retzker}},\ }\href
	{\doibase 10.1103/PhysRevA.85.040302} {\bibfield  {journal} {\bibinfo
			{journal} {Physical Review A}\ }\textbf {\bibinfo {volume} {85}},\ \bibinfo
		{pages} {040302} (\bibinfo {year} {2012})}\BibitemShut {NoStop}%
	\bibitem [{\citenamefont {Cai}\ \emph {et~al.}(2012{\natexlab{a}})\citenamefont
		{Cai}, \citenamefont {Jelezko}, \citenamefont {Katz}, \citenamefont
		{Retzker},\ and\ \citenamefont {Plenio}}]{cai_long-lived_2012}%
	\BibitemOpen
	\bibfield  {author} {\bibinfo {author} {\bibfnamefont {J.}~\bibnamefont
			{Cai}}, \bibinfo {author} {\bibfnamefont {F.}~\bibnamefont {Jelezko}},
		\bibinfo {author} {\bibfnamefont {N.}~\bibnamefont {Katz}}, \bibinfo {author}
		{\bibfnamefont {A.}~\bibnamefont {Retzker}}, \ and\ \bibinfo {author}
		{\bibfnamefont {M.~B.}\ \bibnamefont {Plenio}},\ }\href {\doibase
		10.1088/1367-2630/14/9/093030} {\bibfield  {journal} {\bibinfo  {journal}
			{New Journal of Physics}\ }\textbf {\bibinfo {volume} {14}},\ \bibinfo
		{pages} {093030} (\bibinfo {year} {2012}{\natexlab{a}})}\BibitemShut
	{NoStop}%
	\bibitem [{\citenamefont {Xu}\ \emph {et~al.}(2012)\citenamefont {Xu},
		\citenamefont {Wang}, \citenamefont {Duan}, \citenamefont {Huang},
		\citenamefont {Wang}, \citenamefont {Wang}, \citenamefont {Xu}, \citenamefont
		{Kong}, \citenamefont {Shi}, \citenamefont {Rong},\ and\ \citenamefont
		{Du}}]{xu_coherence-protected_2012}%
	\BibitemOpen
	\bibfield  {author} {\bibinfo {author} {\bibfnamefont {X.}~\bibnamefont
			{Xu}}, \bibinfo {author} {\bibfnamefont {Z.}~\bibnamefont {Wang}}, \bibinfo
		{author} {\bibfnamefont {C.}~\bibnamefont {Duan}}, \bibinfo {author}
		{\bibfnamefont {P.}~\bibnamefont {Huang}}, \bibinfo {author} {\bibfnamefont
			{P.}~\bibnamefont {Wang}}, \bibinfo {author} {\bibfnamefont {Y.}~\bibnamefont
			{Wang}}, \bibinfo {author} {\bibfnamefont {N.}~\bibnamefont {Xu}}, \bibinfo
		{author} {\bibfnamefont {X.}~\bibnamefont {Kong}}, \bibinfo {author}
		{\bibfnamefont {F.}~\bibnamefont {Shi}}, \bibinfo {author} {\bibfnamefont
			{X.}~\bibnamefont {Rong}}, \ and\ \bibinfo {author} {\bibfnamefont
			{J.}~\bibnamefont {Du}},\ }\href {\doibase 10.1103/PhysRevLett.109.070502}
	{\bibfield  {journal} {\bibinfo  {journal} {Physical Review Letters}\
		}\textbf {\bibinfo {volume} {109}},\ \bibinfo {pages} {070502} (\bibinfo
		{year} {2012})}\BibitemShut {NoStop}%
	\bibitem [{\citenamefont {Golter}\ \emph {et~al.}(2014)\citenamefont {Golter},
		\citenamefont {Baldwin},\ and\ \citenamefont
		{Wang}}]{golter_protecting_2014}%
	\BibitemOpen
	\bibfield  {author} {\bibinfo {author} {\bibfnamefont {D.~A.}\ \bibnamefont
			{Golter}}, \bibinfo {author} {\bibfnamefont {T.~K.}\ \bibnamefont {Baldwin}},
		\ and\ \bibinfo {author} {\bibfnamefont {H.}~\bibnamefont {Wang}},\ }\href
	{\doibase 10.1103/PhysRevLett.113.237601} {\bibfield  {journal} {\bibinfo
			{journal} {Physical Review Letters}\ }\textbf {\bibinfo {volume} {113}},\
		\bibinfo {pages} {237601} (\bibinfo {year} {2014})}\BibitemShut {NoStop}%
	\bibitem [{\citenamefont {Trypogeorgos}\ \emph {et~al.}(2018)\citenamefont
		{Trypogeorgos}, \citenamefont {Valdés-Curiel}, \citenamefont {Lundblad},\
		and\ \citenamefont {Spielman}}]{trypogeorgos_synthetic_2018}%
	\BibitemOpen
	\bibfield  {author} {\bibinfo {author} {\bibfnamefont {D.}~\bibnamefont
			{Trypogeorgos}}, \bibinfo {author} {\bibfnamefont {A.}~\bibnamefont
			{Valdés-Curiel}}, \bibinfo {author} {\bibfnamefont {N.}~\bibnamefont
			{Lundblad}}, \ and\ \bibinfo {author} {\bibfnamefont {I.~B.}\ \bibnamefont
			{Spielman}},\ }\href {\doibase 10.1103/PhysRevA.97.013407} {\bibfield
		{journal} {\bibinfo  {journal} {Physical Review A}\ }\textbf {\bibinfo
			{volume} {97}},\ \bibinfo {pages} {013407} (\bibinfo {year}
		{2018})}\BibitemShut {NoStop}%
	\bibitem [{\citenamefont {Mkhitaryan}\ and\ \citenamefont
		{Dobrovitski}(2014)}]{mkhitaryan_decay_2014}%
	\BibitemOpen
	\bibfield  {author} {\bibinfo {author} {\bibfnamefont {V.~V.}\ \bibnamefont
			{Mkhitaryan}}\ and\ \bibinfo {author} {\bibfnamefont {V.~V.}\ \bibnamefont
			{Dobrovitski}},\ }\href {\doibase 10.1103/PhysRevB.89.224402} {\bibfield
		{journal} {\bibinfo  {journal} {Physical Review B}\ }\textbf {\bibinfo
			{volume} {89}},\ \bibinfo {pages} {224402} (\bibinfo {year}
		{2014})}\BibitemShut {NoStop}%
	\bibitem [{\citenamefont {Mkhitaryan}\ \emph {et~al.}(2015)\citenamefont
		{Mkhitaryan}, \citenamefont {Jelezko},\ and\ \citenamefont
		{Dobrovitski}}]{mkhitaryan_highly_2015}%
	\BibitemOpen
	\bibfield  {author} {\bibinfo {author} {\bibfnamefont {V.~V.}\ \bibnamefont
			{Mkhitaryan}}, \bibinfo {author} {\bibfnamefont {F.}~\bibnamefont {Jelezko}},
		\ and\ \bibinfo {author} {\bibfnamefont {V.~V.}\ \bibnamefont
			{Dobrovitski}},\ }\href {\doibase 10.1038/srep15402} {\bibfield  {journal}
		{\bibinfo  {journal} {Scientific Reports}\ }\textbf {\bibinfo {volume} {5}},\
		\bibinfo {pages} {15402} (\bibinfo {year} {2015})}\BibitemShut {NoStop}%
	\bibitem [{\citenamefont {Cai}\ \emph {et~al.}(2012{\natexlab{b}})\citenamefont
		{Cai}, \citenamefont {Naydenov}, \citenamefont {Pfeiffer}, \citenamefont
		{McGuinness}, \citenamefont {Jahnke}, \citenamefont {Jelezko}, \citenamefont
		{Plenio},\ and\ \citenamefont {Retzker}}]{cai_robust_2012}%
	\BibitemOpen
	\bibfield  {author} {\bibinfo {author} {\bibfnamefont {J.-M.}\ \bibnamefont
			{Cai}}, \bibinfo {author} {\bibfnamefont {B.}~\bibnamefont {Naydenov}},
		\bibinfo {author} {\bibfnamefont {R.}~\bibnamefont {Pfeiffer}}, \bibinfo
		{author} {\bibfnamefont {L.~P.}\ \bibnamefont {McGuinness}}, \bibinfo
		{author} {\bibfnamefont {K.~D.}\ \bibnamefont {Jahnke}}, \bibinfo {author}
		{\bibfnamefont {F.}~\bibnamefont {Jelezko}}, \bibinfo {author} {\bibfnamefont
			{M.~B.}\ \bibnamefont {Plenio}}, \ and\ \bibinfo {author} {\bibfnamefont
			{A.}~\bibnamefont {Retzker}},\ }\href {\doibase
		10.1088/1367-2630/14/11/113023} {\bibfield  {journal} {\bibinfo  {journal}
			{New Journal of Physics}\ }\textbf {\bibinfo {volume} {14}},\ \bibinfo
		{pages} {113023} (\bibinfo {year} {2012}{\natexlab{b}})}\BibitemShut
	{NoStop}%
	\bibitem [{\citenamefont {Cohen}\ \emph {et~al.}(2015)\citenamefont {Cohen},
		\citenamefont {Weidt}, \citenamefont {Hensinger},\ and\ \citenamefont
		{Retzker}}]{cohen_multi-qubit_2015}%
	\BibitemOpen
	\bibfield  {author} {\bibinfo {author} {\bibfnamefont {I.}~\bibnamefont
			{Cohen}}, \bibinfo {author} {\bibfnamefont {S.}~\bibnamefont {Weidt}},
		\bibinfo {author} {\bibfnamefont {W.~K.}\ \bibnamefont {Hensinger}}, \ and\
		\bibinfo {author} {\bibfnamefont {A.}~\bibnamefont {Retzker}},\ }\href
	{\doibase 10.1088/1367-2630/17/4/043008} {\bibfield  {journal} {\bibinfo
			{journal} {New Journal of Physics}\ }\textbf {\bibinfo {volume} {17}},\
		\bibinfo {pages} {043008} (\bibinfo {year} {2015})}\BibitemShut {NoStop}%
	\bibitem [{\citenamefont {Teissier}\ \emph {et~al.}(2017)\citenamefont
		{Teissier}, \citenamefont {Barfuss},\ and\ \citenamefont
		{Maletinsky}}]{teissier_hybrid_2017}%
	\BibitemOpen
	\bibfield  {author} {\bibinfo {author} {\bibfnamefont {J.}~\bibnamefont
			{Teissier}}, \bibinfo {author} {\bibfnamefont {A.}~\bibnamefont {Barfuss}}, \
		and\ \bibinfo {author} {\bibfnamefont {P.}~\bibnamefont {Maletinsky}},\
	}\href {\doibase 10.1088/2040-8986/aa5f62} {\bibfield  {journal} {\bibinfo
			{journal} {Journal of Optics}\ }\textbf {\bibinfo {volume} {19}},\ \bibinfo
		{pages} {044003} (\bibinfo {year} {2017})}\BibitemShut {NoStop}%
	\bibitem [{\citenamefont {Cohen}\ \emph {et~al.}(2017)\citenamefont {Cohen},
		\citenamefont {Aharon},\ and\ \citenamefont
		{Retzker}}]{cohen_continuous_2017}%
	\BibitemOpen
	\bibfield  {author} {\bibinfo {author} {\bibfnamefont {I.}~\bibnamefont
			{Cohen}}, \bibinfo {author} {\bibfnamefont {N.}~\bibnamefont {Aharon}}, \
		and\ \bibinfo {author} {\bibfnamefont {A.}~\bibnamefont {Retzker}},\ }\href
	{\doibase 10.1002/prop.201600071} {\bibfield  {journal} {\bibinfo  {journal}
			{Fortschritte der Physik}\ }\textbf {\bibinfo {volume} {65}},\ \bibinfo
		{pages} {1600071} (\bibinfo {year} {2017})}\BibitemShut {NoStop}%
	\bibitem [{\citenamefont {Stark}\ \emph {et~al.}(2017)\citenamefont {Stark},
		\citenamefont {Aharon}, \citenamefont {Unden}, \citenamefont {Louzon},
		\citenamefont {Huck}, \citenamefont {Retzker}, \citenamefont {Andersen},\
		and\ \citenamefont {Jelezko}}]{stark_narrow-bandwidth_2017}%
	\BibitemOpen
	\bibfield  {author} {\bibinfo {author} {\bibfnamefont {A.}~\bibnamefont
			{Stark}}, \bibinfo {author} {\bibfnamefont {N.}~\bibnamefont {Aharon}},
		\bibinfo {author} {\bibfnamefont {T.}~\bibnamefont {Unden}}, \bibinfo
		{author} {\bibfnamefont {D.}~\bibnamefont {Louzon}}, \bibinfo {author}
		{\bibfnamefont {A.}~\bibnamefont {Huck}}, \bibinfo {author} {\bibfnamefont
			{A.}~\bibnamefont {Retzker}}, \bibinfo {author} {\bibfnamefont {U.~L.}\
			\bibnamefont {Andersen}}, \ and\ \bibinfo {author} {\bibfnamefont
			{F.}~\bibnamefont {Jelezko}},\ }\href {\doibase 10.1038/s41467-017-01159-2}
	{\bibfield  {journal} {\bibinfo  {journal} {Nature Communications}\ }\textbf
		{\bibinfo {volume} {8}},\ \bibinfo {pages} {1105} (\bibinfo {year}
		{2017})}\BibitemShut {NoStop}%
	\bibitem [{\citenamefont {Timoney}\ \emph {et~al.}(2011)\citenamefont
		{Timoney}, \citenamefont {Baumgart}, \citenamefont {Johanning}, \citenamefont
		{Varón}, \citenamefont {Plenio}, \citenamefont {Retzker},\ and\
		\citenamefont {Wunderlich}}]{timoney_quantum_2011}%
	\BibitemOpen
	\bibfield  {author} {\bibinfo {author} {\bibfnamefont {N.}~\bibnamefont
			{Timoney}}, \bibinfo {author} {\bibfnamefont {I.}~\bibnamefont {Baumgart}},
		\bibinfo {author} {\bibfnamefont {M.}~\bibnamefont {Johanning}}, \bibinfo
		{author} {\bibfnamefont {A.~F.}\ \bibnamefont {Varón}}, \bibinfo {author}
		{\bibfnamefont {M.~B.}\ \bibnamefont {Plenio}}, \bibinfo {author}
		{\bibfnamefont {A.}~\bibnamefont {Retzker}}, \ and\ \bibinfo {author}
		{\bibfnamefont {C.}~\bibnamefont {Wunderlich}},\ }\href {\doibase
		10.1038/nature10319} {\bibfield  {journal} {\bibinfo  {journal} {Nature}\
		}\textbf {\bibinfo {volume} {476}},\ \bibinfo {pages} {185} (\bibinfo {year}
		{2011})}\BibitemShut {NoStop}%
	\bibitem [{\citenamefont {Aharon}\ \emph {et~al.}(2013)\citenamefont {Aharon},
		\citenamefont {Drewsen},\ and\ \citenamefont
		{Retzker}}]{aharon_general_2013}%
	\BibitemOpen
	\bibfield  {author} {\bibinfo {author} {\bibfnamefont {N.}~\bibnamefont
			{Aharon}}, \bibinfo {author} {\bibfnamefont {M.}~\bibnamefont {Drewsen}}, \
		and\ \bibinfo {author} {\bibfnamefont {A.}~\bibnamefont {Retzker}},\ }\href
	{\doibase 10.1103/PhysRevLett.111.230507} {\bibfield  {journal} {\bibinfo
			{journal} {Physical Review Letters}\ }\textbf {\bibinfo {volume} {111}},\
		\bibinfo {pages} {230507} (\bibinfo {year} {2013})}\BibitemShut {NoStop}%
	\bibitem [{\citenamefont {Aharon}\ \emph {et~al.}(2016)\citenamefont {Aharon},
		\citenamefont {Cohen}, \citenamefont {Jelezko},\ and\ \citenamefont
		{Retzker}}]{aharon_fully_2016}%
	\BibitemOpen
	\bibfield  {author} {\bibinfo {author} {\bibfnamefont {N.}~\bibnamefont
			{Aharon}}, \bibinfo {author} {\bibfnamefont {I.}~\bibnamefont {Cohen}},
		\bibinfo {author} {\bibfnamefont {F.}~\bibnamefont {Jelezko}}, \ and\
		\bibinfo {author} {\bibfnamefont {A.}~\bibnamefont {Retzker}},\ }\href
	{\doibase 10.1088/1367-2630/aa4fd3} {\bibfield  {journal} {\bibinfo
			{journal} {New Journal of Physics}\ }\textbf {\bibinfo {volume} {18}},\
		\bibinfo {pages} {123012} (\bibinfo {year} {2016})}\BibitemShut {NoStop}%
	\bibitem [{\citenamefont {Gruber}\ \emph {et~al.}(1997)\citenamefont {Gruber},
		\citenamefont {Dräbenstedt}, \citenamefont {Tietz}, \citenamefont {Fleury},
		\citenamefont {Wrachtrup},\ and\ \citenamefont
		{Borczyskowski}}]{gruber_scanning_1997}%
	\BibitemOpen
	\bibfield  {author} {\bibinfo {author} {\bibfnamefont {A.}~\bibnamefont
			{Gruber}}, \bibinfo {author} {\bibfnamefont {A.}~\bibnamefont
			{Dräbenstedt}}, \bibinfo {author} {\bibfnamefont {C.}~\bibnamefont {Tietz}},
		\bibinfo {author} {\bibfnamefont {L.}~\bibnamefont {Fleury}}, \bibinfo
		{author} {\bibfnamefont {J.}~\bibnamefont {Wrachtrup}}, \ and\ \bibinfo
		{author} {\bibfnamefont {C.~v.}\ \bibnamefont {Borczyskowski}},\ }\href
	{\doibase 10.1126/science.276.5321.2012} {\bibfield  {journal} {\bibinfo
			{journal} {Science}\ }\textbf {\bibinfo {volume} {276}},\ \bibinfo {pages}
		{2012} (\bibinfo {year} {1997})}\BibitemShut {NoStop}%
	\bibitem [{\citenamefont {Jelezko}\ and\ \citenamefont
		{Wrachtrup}(2006)}]{jelezko_single_2006}%
	\BibitemOpen
	\bibfield  {author} {\bibinfo {author} {\bibfnamefont {F.}~\bibnamefont
			{Jelezko}}\ and\ \bibinfo {author} {\bibfnamefont {J.}~\bibnamefont
			{Wrachtrup}},\ }\href {\doibase 10.1002/pssa.200671403} {\bibfield  {journal}
		{\bibinfo  {journal} {physica status solidi (a)}\ }\textbf {\bibinfo {volume}
			{203}},\ \bibinfo {pages} {3207} (\bibinfo {year} {2006})}\BibitemShut
	{NoStop}%
	\bibitem [{\citenamefont {Maze}\ \emph {et~al.}(2008)\citenamefont {Maze},
		\citenamefont {Stanwix}, \citenamefont {Hodges}, \citenamefont {Hong},
		\citenamefont {Taylor}, \citenamefont {Cappellaro}, \citenamefont {Jiang},
		\citenamefont {Dutt}, \citenamefont {Togan}, \citenamefont {Zibrov},
		\citenamefont {Yacoby}, \citenamefont {Walsworth},\ and\ \citenamefont
		{Lukin}}]{maze_nanoscale_2008}%
	\BibitemOpen
	\bibfield  {author} {\bibinfo {author} {\bibfnamefont {J.~R.}\ \bibnamefont
			{Maze}}, \bibinfo {author} {\bibfnamefont {P.~L.}\ \bibnamefont {Stanwix}},
		\bibinfo {author} {\bibfnamefont {J.~S.}\ \bibnamefont {Hodges}}, \bibinfo
		{author} {\bibfnamefont {S.}~\bibnamefont {Hong}}, \bibinfo {author}
		{\bibfnamefont {J.~M.}\ \bibnamefont {Taylor}}, \bibinfo {author}
		{\bibfnamefont {P.}~\bibnamefont {Cappellaro}}, \bibinfo {author}
		{\bibfnamefont {L.}~\bibnamefont {Jiang}}, \bibinfo {author} {\bibfnamefont
			{M.~V.~G.}\ \bibnamefont {Dutt}}, \bibinfo {author} {\bibfnamefont
			{E.}~\bibnamefont {Togan}}, \bibinfo {author} {\bibfnamefont {A.~S.}\
			\bibnamefont {Zibrov}}, \bibinfo {author} {\bibfnamefont {A.}~\bibnamefont
			{Yacoby}}, \bibinfo {author} {\bibfnamefont {R.~L.}\ \bibnamefont
			{Walsworth}}, \ and\ \bibinfo {author} {\bibfnamefont {M.~D.}\ \bibnamefont
			{Lukin}},\ }\href {\doibase 10.1038/nature07279} {\bibfield  {journal}
		{\bibinfo  {journal} {Nature}\ }\textbf {\bibinfo {volume} {455}},\ \bibinfo
		{pages} {nature07279} (\bibinfo {year} {2008})}\BibitemShut {NoStop}%
	\bibitem [{\citenamefont {Doherty}\ \emph {et~al.}(2013)\citenamefont
		{Doherty}, \citenamefont {Manson}, \citenamefont {Delaney}, \citenamefont
		{Jelezko}, \citenamefont {Wrachtrup},\ and\ \citenamefont
		{Hollenberg}}]{doherty_nitrogen-vacancy_2013}%
	\BibitemOpen
	\bibfield  {author} {\bibinfo {author} {\bibfnamefont {M.~W.}\ \bibnamefont
			{Doherty}}, \bibinfo {author} {\bibfnamefont {N.~B.}\ \bibnamefont {Manson}},
		\bibinfo {author} {\bibfnamefont {P.}~\bibnamefont {Delaney}}, \bibinfo
		{author} {\bibfnamefont {F.}~\bibnamefont {Jelezko}}, \bibinfo {author}
		{\bibfnamefont {J.}~\bibnamefont {Wrachtrup}}, \ and\ \bibinfo {author}
		{\bibfnamefont {L.~C.~L.}\ \bibnamefont {Hollenberg}},\ }\href {\doibase
		10.1016/j.physrep.2013.02.001} {\bibfield  {journal} {\bibinfo  {journal}
			{Physics Reports}\ }\bibinfo {series} {The nitrogen-vacancy colour centre in
			diamond},\ \textbf {\bibinfo {volume} {528}},\ \bibinfo {pages} {1} (\bibinfo
		{year} {2013})}\BibitemShut {NoStop}%
	\bibitem [{\citenamefont {Jacques}\ \emph {et~al.}(2009)\citenamefont
		{Jacques}, \citenamefont {Neumann}, \citenamefont {Beck}, \citenamefont
		{Markham}, \citenamefont {Twitchen}, \citenamefont {Meijer}, \citenamefont
		{Kaiser}, \citenamefont {Balasubramanian}, \citenamefont {Jelezko},\ and\
		\citenamefont {Wrachtrup}}]{jacques_dynamic_2009}%
	\BibitemOpen
	\bibfield  {author} {\bibinfo {author} {\bibfnamefont {V.}~\bibnamefont
			{Jacques}}, \bibinfo {author} {\bibfnamefont {P.}~\bibnamefont {Neumann}},
		\bibinfo {author} {\bibfnamefont {J.}~\bibnamefont {Beck}}, \bibinfo {author}
		{\bibfnamefont {M.}~\bibnamefont {Markham}}, \bibinfo {author} {\bibfnamefont
			{D.}~\bibnamefont {Twitchen}}, \bibinfo {author} {\bibfnamefont
			{J.}~\bibnamefont {Meijer}}, \bibinfo {author} {\bibfnamefont
			{F.}~\bibnamefont {Kaiser}}, \bibinfo {author} {\bibfnamefont
			{G.}~\bibnamefont {Balasubramanian}}, \bibinfo {author} {\bibfnamefont
			{F.}~\bibnamefont {Jelezko}}, \ and\ \bibinfo {author} {\bibfnamefont
			{J.}~\bibnamefont {Wrachtrup}},\ }\href {\doibase
		10.1103/PhysRevLett.102.057403} {\bibfield  {journal} {\bibinfo  {journal}
			{Physical Review Letters}\ }\textbf {\bibinfo {volume} {102}},\ \bibinfo
		{pages} {057403} (\bibinfo {year} {2009})}\BibitemShut {NoStop}%
	\bibitem [{\citenamefont {Ivády}\ \emph {et~al.}(2015)\citenamefont {Ivády},
		\citenamefont {Szász}, \citenamefont {Falk}, \citenamefont {Klimov},
		\citenamefont {Christle}, \citenamefont {Janzén}, \citenamefont {Abrikosov},
		\citenamefont {Awschalom},\ and\ \citenamefont
		{Gali}}]{ivady_theoretical_2015}%
	\BibitemOpen
	\bibfield  {author} {\bibinfo {author} {\bibfnamefont {V.}~\bibnamefont
			{Ivády}}, \bibinfo {author} {\bibfnamefont {K.}~\bibnamefont {Szász}},
		\bibinfo {author} {\bibfnamefont {A.~L.}\ \bibnamefont {Falk}}, \bibinfo
		{author} {\bibfnamefont {P.~V.}\ \bibnamefont {Klimov}}, \bibinfo {author}
		{\bibfnamefont {D.~J.}\ \bibnamefont {Christle}}, \bibinfo {author}
		{\bibfnamefont {E.}~\bibnamefont {Janzén}}, \bibinfo {author} {\bibfnamefont
			{I.~A.}\ \bibnamefont {Abrikosov}}, \bibinfo {author} {\bibfnamefont {D.~D.}\
			\bibnamefont {Awschalom}}, \ and\ \bibinfo {author} {\bibfnamefont
			{A.}~\bibnamefont {Gali}},\ }\href {\doibase 10.1103/PhysRevB.92.115206}
	{\bibfield  {journal} {\bibinfo  {journal} {Physical Review B}\ }\textbf
		{\bibinfo {volume} {92}},\ \bibinfo {pages} {115206} (\bibinfo {year}
		{2015})}\BibitemShut {NoStop}%
	\bibitem [{\citenamefont {Cywinski}\ \emph {et~al.}(2008)\citenamefont
		{Cywinski}, \citenamefont {Lutchyn}, \citenamefont {Nave},\ and\
		\citenamefont {Das~Sarma}}]{cywinski_how_2008}%
	\BibitemOpen
	\bibfield  {author} {\bibinfo {author} {\bibfnamefont {L.}~\bibnamefont
			{Cywinski}}, \bibinfo {author} {\bibfnamefont {R.~M.}\ \bibnamefont
			{Lutchyn}}, \bibinfo {author} {\bibfnamefont {C.~P.}\ \bibnamefont {Nave}}, \
		and\ \bibinfo {author} {\bibfnamefont {S.}~\bibnamefont {Das~Sarma}},\ }\href
	{\doibase 10.1103/PhysRevB.77.174509} {\bibfield  {journal} {\bibinfo
			{journal} {Physical Review B}\ }\textbf {\bibinfo {volume} {77}},\ \bibinfo
		{pages} {174509} (\bibinfo {year} {2008})}\BibitemShut {NoStop}%
	\bibitem [{\citenamefont {Wang}\ and\ \citenamefont
		{Takahashi}(2013)}]{wang_spin_2013}%
	\BibitemOpen
	\bibfield  {author} {\bibinfo {author} {\bibfnamefont {Z.-H.}\ \bibnamefont
			{Wang}}\ and\ \bibinfo {author} {\bibfnamefont {S.}~\bibnamefont
			{Takahashi}},\ }\href {\doibase 10.1103/PhysRevB.87.115122} {\bibfield
		{journal} {\bibinfo  {journal} {Physical Review B}\ }\textbf {\bibinfo
			{volume} {87}},\ \bibinfo {pages} {115122} (\bibinfo {year}
		{2013})}\BibitemShut {NoStop}%
	\bibitem [{\citenamefont {Binder}\ \emph {et~al.}(2017)\citenamefont {Binder},
		\citenamefont {Stark}, \citenamefont {Tomek}, \citenamefont {Scheuer},
		\citenamefont {Frank}, \citenamefont {Jahnke}, \citenamefont {Müller},
		\citenamefont {Schmitt}, \citenamefont {Metsch}, \citenamefont {Unden},
		\citenamefont {Gehring}, \citenamefont {Huck}, \citenamefont {Andersen},
		\citenamefont {Rogers},\ and\ \citenamefont {Jelezko}}]{binder_qudi_2017}%
	\BibitemOpen
	\bibfield  {author} {\bibinfo {author} {\bibfnamefont {J.~M.}\ \bibnamefont
			{Binder}}, \bibinfo {author} {\bibfnamefont {A.}~\bibnamefont {Stark}},
		\bibinfo {author} {\bibfnamefont {N.}~\bibnamefont {Tomek}}, \bibinfo
		{author} {\bibfnamefont {J.}~\bibnamefont {Scheuer}}, \bibinfo {author}
		{\bibfnamefont {F.}~\bibnamefont {Frank}}, \bibinfo {author} {\bibfnamefont
			{K.~D.}\ \bibnamefont {Jahnke}}, \bibinfo {author} {\bibfnamefont
			{C.}~\bibnamefont {Müller}}, \bibinfo {author} {\bibfnamefont
			{S.}~\bibnamefont {Schmitt}}, \bibinfo {author} {\bibfnamefont {M.~H.}\
			\bibnamefont {Metsch}}, \bibinfo {author} {\bibfnamefont {T.}~\bibnamefont
			{Unden}}, \bibinfo {author} {\bibfnamefont {T.}~\bibnamefont {Gehring}},
		\bibinfo {author} {\bibfnamefont {A.}~\bibnamefont {Huck}}, \bibinfo {author}
		{\bibfnamefont {U.~L.}\ \bibnamefont {Andersen}}, \bibinfo {author}
		{\bibfnamefont {L.~J.}\ \bibnamefont {Rogers}}, \ and\ \bibinfo {author}
		{\bibfnamefont {F.}~\bibnamefont {Jelezko}},\ }\href {\doibase
		10.1016/j.softx.2017.02.001} {\bibfield  {journal} {\bibinfo  {journal}
			{SoftwareX}\ }\textbf {\bibinfo {volume} {6}},\ \bibinfo {pages} {85}
		(\bibinfo {year} {2017})}\BibitemShut {NoStop}%
\end{thebibliography}
\end{document}